\documentclass[aps,pra,onecolumn,floatfix]{revtex4}

\usepackage{graphicx}
\usepackage{amsmath,amssymb}
\usepackage{graphpap}

\usepackage{amsmath, amsthm, amssymb}
\usepackage{graphicx,color,psfrag}
\usepackage[normalem]{ulem}
\usepackage{thmtools,mathtools}

\DeclareMathAlphabet{\gcal}{OMS}{cmsy}{m}{n}

\newcommand{\Lfun}[2]{{\gcal L}^{(#1)}_{#2}}
\newcommand{\Tr}[1]{\operatorname{Tr} \left\{ #1 \right\}}

\DeclarePairedDelimiter\abs{\lvert}{\rvert}%
\DeclarePairedDelimiter\norm{\lVert}{\rVert}%
\makeatletter
\let\oldabs\abs
\def\abs{\@ifstar{\oldabs}{\oldabs*}}
\let\oldnorm\norm
\def\norm{\@ifstar{\oldnorm}{\oldnorm*}}
\makeatother

\begin{document}

\title{Local temperature of an interacting quantum system far from equilibrium}

\author{
Charles A.\ Stafford
}

\affiliation{Department of Physics, University of Arizona, 1118 E.\ 4th St., Tucson, AZ, 85721}

\date{\today}

\begin{abstract}
A theory of local temperature measurement of an interacting quantum electron system far from equilibrium via a floating thermoelectric probe 
is developed.  It is shown that the local temperature so defined is consistent with the zeroth, first, second, and third laws of thermodynamics,
provided the probe-system coupling is weak and broad band. % (ideal temperature measurement).  
For non-broad-band probes,
%general probe-system couplings, 
the local temperature obeys the Clausius form of the second law 
and the third law exactly, but there are corrections to the zeroth and first laws that are higher-order in the Sommerfeld expansion. 
The corrections to the zeroth and first laws are related, and can be interpreted in terms of the error of a nonideal temperature measurement.
These results also hold for systems at negative absolute temperature.
\end{abstract}


\maketitle

\section{Introduction}
\label{sec:intro}

Nonequilibrium Green's functions (NEGF) \cite{Stefanucci13} 
provide a systematic method to study the local properties of interacting quantum systems far from equilibrium; however, a corresponding
thermodynamic description has generally been lacking, outside the limit where a local equilibrium exists \cite{Groo62}.  
The concept of a local temperature 
has been extended to nonequilibrium
systems under the assumption of local equilibrium \cite{Groo62}, but it has proven far more challenging to
generalize 
to systems where the local equilibrium hypothesis does not
hold \cite{Cug11,Cas03}. Without local equilibrium,  different temperatures may be obtained by different
measurement protocols~\cite{Cas03}.  
Furthermore, out of equilibrium, the temperature distributions of different microscopic
degrees of freedom (e.g., electrons, phonons, nuclear spins) do not, in general, coincide, so that one has to distinguish between measurements of the 
electron temperature \cite{Engquist81,Dubi09c}, the lattice temperature \cite{Ming10,Galperin07}, the nuclear temperature \cite{NMRbook}, etc.
This distinction is particularly acute in the 
extreme limit of elastic quantum transport \cite{Bergfield13demon,Meair14,Bergfield15}, where electron and phonon temperatures are completely decoupled.
The consensus has thus been that 
the various schemes for measuring the temperature of a system far from equilibrium
can at best deliver an operational definition of the local temperature.

In this article, we systematically re-examine this fundamental issue, building upon the findings of Meair et al.\ \cite{Meair14}, 
%While the local temperature $\bar{T}_p$ of a nonequilibrium quantum system measured by a floating thermoelectric probe 
%Meair et al.\
who
argued that the temperature $\bar{T}_p$ measured by a floating thermoelectric probe can be interpreted as the
local temperature of a nonequilibrium electron system, consistent with the laws of thermodynamics.  
While $\bar{T}_p$ %so defined 
does not have the same fundamental basis in statistical mechanics as the temperature of an equilibrium system,
nonetheless
%In particular, 
it was shown that
(i) $\bar{T}_p$ is largely independent of the details of the probe-sample coupling; (ii) the temperature inferred from an independent
electrical noise
measurement coincides with that measured by a floating probe; and (iii) the temperature so defined is consistent with both the zeroth and
second laws of thermodynamics.  These results were obtained within linear response and to leading order in the Sommerfeld expansion.  
%This definition of a local electron temperature, being consistent with many of the conditions satisfied by the 
%temperature of a thermodynamic system at equilibrium, goes well beyond the operational definitions previously proposed.

In the present article, a theory of local temperature measurement is developed that
extends the analysis of Ref.\ \cite{Meair14} to interacting quantum systems 
under steady-state conditions arbitrarily far from equilibrium, using the method of NEGF. 
In addition to the zeroth and second laws, the conditions under which $\bar{T}_p$ 
is consistent with the first and third laws of thermodynamics are investigated.
It is shown that the local temperature defined by a floating thermoelectric probe is consistent with the zeroth, first, second, and third laws of
thermodynamics, provided the probe-system coupling is weak and broad-band.
For non-broad-band probes, %general probe-system couplings, 
the local temperature obeys the Clausius form of the second law
and the third law exactly, but there are corrections to the zeroth and first laws that are higher-order in the Sommerfeld expansion.  
The exact agreement with Clausius's statement of the second law and with the third law implies that the local temperature metric
$\bar{T}_p$ defines an ordering of temperatures, and an absolute zero, but not necessarily an absolute temperature scale.  
The corrections to the zeroth and first laws are shown to be related, and can be interpreted in terms of the error of a nonideal temperature measurement.
This analysis also applies to systems with negative absolute temperature \cite{Ramsey56,Braun13,Carr13,Shastry16} (population inversion).

It is also shown that for a probe with broad-band coupling, 
$\bar{T}_p$ is directly related to the mean local excitation energy of the system, in the same way as it is in an 
equilibrium system.  Our findings make a compelling case to interpret the temperature measured by a noninvasive, broad-band thermoelectric probe as
the local temperature of a nonequilibrium electron system.  This definition goes far beyond a mere operational notion, although it does not have the
same fundamental status as the temperature of a system in equilibrium.

%The first law defines absolute
%temperature differences, and it was shown that, at least within linear response,
%discrepancies with the first law arise from deviations from ideal measurement (zeroth law).

This article is organized as follows:  In Sec.\ \ref{sec:intro_NEGF}, the probe equilibration conditions are defined and expressed using the NEGF formalism. 
Prior results \cite{Bergfield13demon,Bergfield15} obtained within linear-response theory are also summarized here, introducing the Onsager coefficients 
that are useful in the sequel.  In Sec.\ \ref{sec:local}, the local spectrum and distribution function sampled by the probe are defined,
and the charge and heat currents flowing into the probe are related to these local properties of the nonequilibrium quantum system.
It is shown that the probe equilibration problem is determined entirely by the local occupancy and energy of the system for a noninvasive, broad-band
probe.
Secs.\ \ref{sec:0th_law}--\ref{sec:3rd_law} examine the extent to which the temperature $\bar{T}_p$ measured by a noninvasive local probe is consistent with
the 0th, 1st, 2nd, and 3rd laws of thermodynamics, respectively, even when the system probed is arbitrarily far from equilibrium.
Sec.\ \ref{sec:conclusions} draws together the various threads presented throughout the paper, and concludes 
that $\bar{T}_p$ provides far more than a mere operational definition of
temperature out of equilibrium.  Appendix \ref{sec:AppA} derives some important properties of the local spectrum and distribution function of a nonequilibrium
steady-state, while Appendix \ref{sec:appendixB} examines an alternative to the zeroth-law scenario investigated in Sec.\ \ref{sec:0th_law}.

\section{Current formula and probe temperature}
\label{sec:intro_NEGF}

Our approach is motivated by the experimental technique of
scanning thermal microscopy \cite{Majumdar99}, whose resolution has recently been brought down to
the nanometer range 
\cite{Kim11,Yu11,Kim12,Fabian12}.
The system's local temperature is defined via an external local probe  weakly coupled
to the system via a tunnel barrier~\cite{Bergfield13demon}.
At its other end, the probe is connected to a macroscopic electron
reservoir whose chemical potential and temperature ``float'' until neither electric current nor heat current flow
between the probe and the system
\cite{Bergfield13demon,Meair14}:
\begin{equation}
I_p^{(\nu)} =0, \; \nu=0,1,
\label{eq:def_probe}
\end{equation}
where $-eI_p^{(0)}$ and $I_p^{(1)}$ are the electric current and heat current, respectively, flowing into the probe.
The probe is then in local equilibrium with a system that is itself arbitrarily far from equilibrium.
Several investigators have proposed related definitions of a floating thermal probe
\cite{Engquist81,Jacquet09,Dubi09a,Dubi09b,Dubi09c,Caso11,Galperin11,Jacquet11,Sanchez11,Caso12,Ye15}. 
It should be noted that a number of other schemes for measuring the temperature of electron systems also exist \cite{Giazotto06}, but
these generally lack the high spatial resolution available in a scanning probe.

The starting point of our analysis is the NEGF formula 
for the steady-state electric 
and heat 
currents flowing into a probe coupled locally to a nonequilibrium quantum system with arbitrary interactions \cite{Meir92,Bergfield09b}
\begin{equation}
I_p^{(\nu)}= - \frac{i}{h} \int_{-\infty}^\infty\!\! d\omega (\omega-\mu_p)^\nu \,
{\rm Tr}\left\{ \Gamma^p(\omega)\left( G^<(\omega) + f_p(\omega)\left[G^r(\omega)\!-\!G^a(\omega)\right]\right) \right\},
\label{eq:both_currents}
\end{equation}
where $\nu=1$ gives the electronic contribution to the heat current and $\nu=0$ the electron number current.
The probe is assumed to consist of a noninteracting electron reservoir
with Fermi-Dirac distribution 
\begin{equation}
f_p(\omega)=\{1+\exp[(\omega-\mu_p)/k_B T_p]\}^{-1}
\end{equation}
and tunneling-width matrix
\begin{equation}
\left[\Gamma^{p}(\omega)\right]_{n\sigma,m\sigma'}=2\pi\delta_{\sigma\sigma'} \sum_{k\in p}V_{nk} V_{mk}^* \, \delta(\omega-\epsilon_{k\sigma}).
\end{equation}
Here $|n\rangle$, $|m\rangle$ are single-particle basis orbitals (e.g., atomic orbitals) in the system, while the states in the probe are labeled
$|k\rangle$.
The coupling matrix elements $V_{nk}$ can be calculated in the tunneling regime using standard methods for scanning probes
\cite{Chen93,Bergfield13demon}.
In Eq.\ (\ref{eq:both_currents}),
$G^r(\omega)$, $G^a(\omega)$, and $G^<(\omega)$ are Fourier transforms of the retarded, advanced, and Keldysh ``lesser'' Green's functions describing
electron propagation/occupancy within the system \cite{Stefanucci13}:
\begin{equation}
G^r_{n\sigma,m\sigma'}(t)=-i\theta(t)\langle \{d_{n\sigma}(t),d_{m\sigma'}^\dagger(0)\}\rangle,
\end{equation}
\begin{equation}
G^a_{n\sigma,m\sigma'}(t)=i\theta(-t)\langle \{d_{n\sigma}(t),d_{m\sigma'}^\dagger(0)\}\rangle,
\end{equation}
and 
\begin{equation}
G^<_{n\sigma,m\sigma'}(t)=i\langle d_{m\sigma'}^\dagger(0)\, d_{n\sigma}(t) \rangle, 
\end{equation}
respectively.
Eq.\ (\ref{eq:both_currents}) is an exact formal result, 
valid for arbitrary interactions and for arbitrary steady-state thermal and/or electric bias.

The probe temperature of an interacting electron system with arbitrary bias is determined by solving the conditions
(\ref{eq:def_probe})
with $I_p^{(\nu)}$ given by Eq.\ (\ref{eq:both_currents}).
%Eq.\ (\ref{eq:def_probe}) gives the conditions under which the probe is in local equilibrium with the sample (which is itself arbitrarily far from equilibrium).
Eqs.\ (\ref{eq:def_probe}) and (\ref{eq:both_currents}) represent two coupled nonlinear equations for the two unknowns, $T_p$ and $\mu_p$.  
{\it A priori}, a solution to the probe equilibration problem might not exist at all, or might not be unique if it did exist. However, it was shown in 
Ref.\ \cite{Shastry16} that for any weak probe-system coupling $\Gamma^{p}(\omega)$, the solution to Eqs.\ (\ref{eq:def_probe}) and (\ref{eq:both_currents})
exists and is unique.  $T_p$ was shown to be positive provided the system does not have local population inversion, and negative if it does.

%With this foundational issue settled, 
Given that the probe equilibration conditions (\ref{eq:def_probe}) constitute a well-posed problem \cite{Shastry16},
the present article addresses the related question of how the measured value of $T_p$ depends
on the actual probe-system coupling $\Gamma^{p}(\omega)$ for a system far from equilibrium. 
That is, to what extent do {\em various thermometers measure different temperatures} of the same
nonequilibrium quantum system?  The thermodynamic interpretation of such a local nonequilibrium temperature is also explored.

\subsection{Linear response results}

We summarize here the formalism for linear thermoelectric response of an open quantum system, because several of the key concepts and
formulas will be useful by analogy in treating the far from equilibrium system.
Consider a general system
with $M$ electrical contacts.  Each contact $\alpha$ is connected to a reservoir at temperature $T_\alpha$
and electrochemical potential $\mu_\alpha$.  
In linear response, the electric current $-eI^{(0)}_\alpha$  and heat current $I^{(1)}_\alpha$ flowing into reservoir $\alpha$ may be expressed as
\begin{equation}
        I_\alpha^{(\nu)} = \sum_{\beta=1}^M \left[ \Lfun{\nu}{\alpha\beta} (\mu_\beta-\mu_\alpha) 
+ \Lfun{\nu+1}{\alpha\beta}\left(\frac{T_\beta-T_\alpha}{T_0}\right) \right],
        \label{eq:LinearResponse_Currents}
\end{equation}
where $\Lfun{\nu}{\alpha\beta}$ ($\nu=0,1,2$) is an Onsager linear-response coefficient \cite{Onsager31}.  

In a thermal transport experiment, the system is driven out of equilibrium by a thermal bias applied between the hot and cold electrodes,
but the system forms an open electric circuit.  Under these conditions, the chemical potentials $\mu_\alpha$ may be
eliminated from Eq.\ (\ref{eq:LinearResponse_Currents}), yielding the following expression for the total heat current flowing into the probe,
which forms the third terminal of the thermoelectric circuit:
\begin{equation}
I_p^Q \equiv I_p^{(1)}
=\sum_{\beta=1}^2 \tilde{\kappa}_{p\beta} (T_\beta - T_p) + \kappa_{p0}(T_0-T_p).
\label{eq:probe_I1}
\end{equation}
Here $\tilde{\kappa}_{\alpha\beta}$ is the 
thermal conductance between electrodes $\alpha$ and $\beta$, and $\kappa_{p0}$ is the thermal
coupling of the probe to the ambient environment at temperature $T_0$.
The environment
could be, for example, the black-body radiation or
gaseous atmosphere surrounding the circuit, or the
cantilever/driver on which the temperature probe
is mounted. 

Eqs.\ (\ref{eq:def_probe}) and (\ref{eq:probe_I1}) can be solved for the temperature of a probe
in thermal and electrical equilibrium with, and coupled locally to the system 
\cite{Bergfield13demon}
\begin{equation}
\bar{T}_p=\frac{\tilde{\kappa}_{p1} T_1 + \tilde{\kappa}_{p2} T_2 + \kappa_{p0} T_0}{\tilde{\kappa}_{p1} + \tilde{\kappa}_{p2}+  \kappa_{p0}}.
\label{eq:Tp}
\end{equation}
The effect of 
$\kappa_{p0}$ 
on local temperature measurement,
an important issue in 
nanoscale thermometry
\cite{Kim11,Yu11,Kim12,Fabian12},
was discussed in Refs.\ \onlinecite{Bergfield13demon,Bergfield15}.
In the present manuscript, we are concerned with establishing the fundamental theoretical basis for defining a local temperature of a nonequilibrium quantum 
system, not with the nonidealities inherent in experimental thermometry.
Therefore, unless otherwise specified, we will take $\kappa_{p0}=0$ in the remainder of the present manuscript.

In the absence of an external magnetic field $\Lfun{\nu}{\alpha\beta} =  \Lfun{\nu}{\beta\alpha}$ 
and the 
thermal conductances 
are given by \cite{Bergfield13demon}
\begin{equation}
\tilde{\kappa}_{\alpha\beta} = \frac{1}{T}\left[\Lfun{2}{\alpha\beta} 
-\frac{\left[\Lfun{1}{\alpha\beta}\right]^2}{\tilde{\gcal L}_{\alpha\beta}^{(0)}}
- {\gcal L}^{(0)} \!
\left(
\frac{\Lfun{1}{\alpha\gamma}\Lfun{1}{\alpha\beta}}{\Lfun{0}{\alpha\gamma}\Lfun{0}{\alpha\beta}}
+\frac{\Lfun{1}{\gamma\beta}\Lfun{1}{\alpha\beta}}{\Lfun{0}{\gamma\beta}\Lfun{0}{\alpha\beta}}
-\frac{
\Lfun{1}{\alpha\gamma}\Lfun{1}{\gamma\beta}
}{
\Lfun{0}{\alpha\gamma}\Lfun{0}{\gamma\beta}
}
\right)
\right], 
\label{eq:kappatilde}
\end{equation}
with
\begin{equation}
\tilde{\gcal L}_{\alpha\beta}^{(0)}=    \Lfun{0}{\alpha\beta}+
\frac{\Lfun{0}{\alpha\gamma}\Lfun{0}{\gamma\beta}}{\Lfun{0}{\alpha\gamma}+\Lfun{0}{\gamma\beta}}
\label{eq:three_term_L0}
\end{equation}
and 
\begin{equation}
\frac{1}{{\gcal L}^{(0)}} = \frac{1}{\Lfun{0}{12}} + \frac{1}{\Lfun{0}{13}} + \frac{1}{\Lfun{0}{23}}.
\label{eq:L0_series}
\end{equation}

\subsection{Elastic transport}

Within elastic electron transport theory, 
the linear response coefficients needed to evaluate Eq.\ (\ref{eq:Tp}) are given by
\cite{Sivan86,Bergfield09b,Bergfield10a}
\begin{equation}
\Lfun{\nu}{\alpha\beta}
= \frac{1}{h} \int d\omega \; (\omega-\mu_0)^{\nu}\,T_{\alpha\beta}(\omega) \left(-\frac{\partial f_0}{\partial \omega}\right),
\label{eq:Lnu}  
\end{equation}
where $f_0$ is the equilibrium Fermi-Dirac distribution of the electrodes at chemical potential $\mu_0$ and temperature $T_0$.
The elastic transmission function may be expressed as \cite{Datta95,Bergfield09a}
\begin{equation}
T_{\alpha\beta}(\omega)={\rm Tr}\left\{ \Gamma^\alpha(\omega) G^r(\omega) \Gamma^\beta(\omega) G^a(\omega)\right\},
\label{eq:transmission_prob}
\end{equation}
where $\Gamma^\alpha(\omega)$ is the tunneling-width matrix for lead $\alpha$.

\section{Relation of probe currents to local properties of the system}
\label{sec:local}

\subsection{Local properties of the nonequilibrium system}

One can define the {\em mean local spectrum} sampled by the probe as
\begin{equation}
\bar{A}(\omega) \equiv {\rm Tr}\left\{\Gamma^p(\omega) A(\omega)\right\} /{\rm Tr} \left\{\Gamma^p(\omega) \right\},
\label{eq:Abar}
\end{equation}
where the spectral function of the (nonequilibrium) system is
\begin{equation}
A(\omega) =\frac{i}{2\pi} \left(G^r(\omega)- G^a(\omega)\right).
\label{eq:spectrum}
\end{equation}
Eq.\ (\ref{eq:Abar}) defines a density of states averaged over the orbitals coupled to the probe.  In the tunneling regime, the probe-sample
coupling decreases exponentially with distance, so $\bar{A}(\omega)$ is a measure of the local density of states.

In equilibrium, $G^<$ may be expressed as
\begin{equation}
G^<_{\rm eq}(\omega) = 2\pi i A(\omega) f_{\rm eq}(\omega).
\label{eq:Geq<}
\end{equation}
This relation motivates the following definition of the {\em local nonequilibrium distribution function}, as sampled by the probe
\begin{equation}
f_s(\omega) \equiv 
\frac{{\rm Tr}\left\{\Gamma^p(\omega) G^<(\omega)\right\} }{2\pi i \, {\rm Tr} \left\{\Gamma^p(\omega) A(\omega)\right\} }.
\label{eq:fs}
\end{equation}
$\bar{A}(\omega)$ and $f_s(\omega)$ satisfy the necessary conditions for a spectrum and a distribution function, respectively.  In particular,
$\bar{A}(\omega)\geq 0$ and $0\leq f_s(\omega)\leq 1$ (see Appendix \ref{sec:AppA} and Ref.\ \cite{Shastry16} for proofs and further discussion).

The mean occupancy and energy of the electronic orbitals sampled by the probe are
\begin{eqnarray}
\langle N \rangle & \equiv & 
\int \frac{d\omega}{2\pi i} \frac{{\rm Tr} \left\{\Gamma^p(\omega) G^<(\omega)\right\} }{ {\rm Tr} \left\{\Gamma^p(\omega)\right\}}
= \int_{-\infty}^\infty\!\! d\omega \, \bar{A}(\omega) f_s(\omega),
\label{eq:local_N}\\
\langle E \rangle & \equiv & 
\int \frac{d\omega}{2\pi i} \frac{ \omega \,{\rm Tr} \left\{\Gamma^p(\omega) G^<(\omega)\right\} }{ {\rm Tr} \left\{\Gamma^p(\omega)\right\}}
= \int_{-\infty}^\infty\!\! d\omega \, \omega \bar{A}(\omega) f_s(\omega),
\label{eq:local_E}
\end{eqnarray}
respectively.

For the case of maximally local coupling of the probe to the system
\begin{equation}
\left[\Gamma^p(\omega)\right]_{ij} = \bar{\Gamma}^p(\omega) \delta_{in} \delta_{jn},
\label{eq:Gamma_loc}
\end{equation}
where $n$ is a single localized orbital in the sample,
$\bar{A}(\omega)=A_{nn}(\omega)\equiv \rho(\omega)$ 
is just the local density of states and
\begin{equation}
f_s(\omega) = f_n(\omega) = \frac{G^<_{nn}(\omega)}{G^<_{nn}(\omega)-G^>_{nn}(\omega)}.
\label{eq:fn}
\end{equation}

\subsubsection{Elastic transport regime} 
\label{sec:elastic}

In the regime of elastic quantum transport, one can express 
\begin{equation}
f_s(\omega) = \sum_{\alpha=1}^M \lambda_\alpha (\omega) f_\alpha(\omega),
\label{eq:injectivity}
\end{equation}
where 
$f_\alpha(\omega)$ is the equilibrium Fermi-Dirac distribution of reservoir $\alpha$ and
$\lambda_\alpha(\omega)=\rho_\alpha(\omega)/\rho(\omega)$, where
\begin{equation}
\rho_\alpha (\omega)= \frac{1}{2\pi}\left[G^r(\omega)\Gamma^\alpha(\omega)G^a(\omega)\right]_{nn}
\end{equation}
is the {\em injectivity} \cite{lpdos,Gramespacher97} of reservoir $\alpha$, i.e., the local density
of states associated with electrons injected by $\alpha$. The coefficients $\lambda_\alpha(\omega)$ satisfy the condition
\begin{equation}
1=\sum_\alpha \lambda_\alpha(\omega).
\end{equation}
In the elastic transport regime, 
the local nonequilibrium distribution function $f_s(\omega)$ is thus a linear combination of the various Fermi functions of the reservoirs, with
energy-dependent coefficients.  For a quantum system connected to source and drain electrodes under electrical bias, this leads to an energy 
distribution with two characteristic steps at the source and drain Fermi energies (see Fig.\ \ref{fig:local_energy}), 
as observed experimentally in mesoscopic metal wires
\cite{Pothier97a,Pothier97b}.  
For a fermi system, the coefficients $\lambda_\alpha(\mu_0)$ exhibit 
characteristic $2k_F$ oscillations as a function of position \cite{lpdos,Gramespacher97}, leading to oscillations of the local energy density and temperature
\cite{Dubi09c,Bergfield13demon,Meair14,Bergfield15} in the linear-response regime.

\subsection{Effective two-terminal current formulas}

It is useful to rewrite Eq.\ (\ref{eq:both_currents}) in terms of the local distribution $f_s(\omega)$ within the system, as sampled by the probe.  
Using Eqs.\ (\ref{eq:spectrum}) and (\ref{eq:fs}), Eq.\ (\ref{eq:both_currents}) can be rewritten as
\begin{equation}
I_p^{(\nu)}= \frac{1}{\hbar} \int_{-\infty}^\infty\!\! d\omega (\omega-\mu_p)^\nu \,
{\rm Tr}\left\{ \Gamma^p(\omega) A(\omega)\right\} [f_s(\omega) -f_p(\omega)].
\label{eq:Ip_fs}
\end{equation}
This has the structure of a two-terminal current formula with sample-probe ``transmission function'' 
\begin{equation}
T_{ps}(\omega) = 2\pi\,{\rm Tr} \left\{ \Gamma^p(\omega) A(\omega)\right\}.  
\label{eq:T_ps}
\end{equation}
Note, however, that there is no assumption of elastic transport, and
that $f_s$ is not in general an equilibrium distribution.

For a given bias of the system, let us {\em denote the Fermi-Dirac distribution of the probe once it has reached local equilibrium with the system} [as 
defined by Eq.\ (\ref{eq:def_probe})] {\em by} $\bar{f}_p(\omega)$.  Let us now derive a formula for the currents into the probe when the probe is biased away
from this local equilibrium point.  
%For simplicity, we will assume that the local nonequilibrium distribution $f_s(\omega)$ is independent of 
%the probe bias.  There is nothing fundamental about this assumption, but it will make the following analysis and that in Sec.\ 
%\ref{sec:2nd_law} clearer conceptually.  
We will assume that the local nonequilibrium distribution $f_s(\omega)$ is independent of 
the probe bias (non-invasive probe; see Ref.\ \cite{Shastry16}).  
%There is nothing fundamental about this assumption, but it will make the following analysis and that in Sec.\ \ref{sec:2nd_law} clearer conceptually.  
We note that
$f_s(\omega)$ is independent of probe bias provided $\Gamma^p \ll \Gamma^\alpha$ $\forall \alpha \neq p$ (weak probe-sample coupling).
Writing $f_s -f_p= f_s-\bar{f}_p +\bar{f}_p -f_p$, it is evident that $I_p^{(0)}$ is given by Eq.\ (\ref{eq:Ip_fs})
with the local nonequilibrium distribution $f_s$ replaced by the equilibrium distribution $\bar{f}_p$.
To see that the same holds for $I_p^{(1)}$, one can also write $\omega-\mu_p$ in the integrand of Eq.\ (\ref{eq:Ip_fs}) as
$\omega-\bar{\mu}_p+\bar{\mu}_p-\mu_p$, and note that all the integrals involving $f_s-\bar{f}_p$ vanish due to the conditions of Eq.\ (\ref{eq:def_probe}).
The currents flowing into the probe are thus given exactly by the effective two-terminal formula
\begin{equation}
I_p^{(\nu)}= \frac{1}{h} \int_{-\infty}^\infty\!\! d\omega (\omega-\mu_p)^\nu \,
T_{ps}(\omega) [\bar{f}_p(\omega) -f_p(\omega)],
\label{eq:Ip_2term}
\end{equation}
where both $f_p$ and $\bar{f}_p$ are equilibrium Fermi-Dirac distributions.
Although this formula has the same form as the two-terminal current formula of elastic transport theory \cite{Fisher81,Buttiker86,Sivan86,Bergfield09b}, 
note that it holds for arbitrary interactions within the system and that no assumption of elastic transport has been made.
These effects are encoded in the spectral function $A(\omega)$ of the interacting nonequilibrium system appearing in Eq.\ (\ref{eq:T_ps}).

Because Eq.\ (\ref{eq:Ip_2term}) has the same form as the two-terminal current formula of elastic transport theory, 
one can define Onsager coefficients for the probe-sample junction analogous to Eq.\ (\ref{eq:Lnu}):
\begin{equation}
\Lfun{\nu}{ps}
= \frac{1}{h} \int d\omega \; (\omega-\bar{\mu}_p)^{\nu}\,T_{ps}(\omega) \left(-\frac{\partial \bar{f}_p}{\partial \omega}\right).
\label{eq:Lnu_ps}  
\end{equation}
In terms of these coefficients, one may express the thermopower and thermal conductance of the probe-sample junction as
\begin{equation}
S_{ps} = -\frac{1}{eT_p} \frac{\Lfun{1}{ps}}{\Lfun{0}{ps}},
\label{eq:S_ps}
\end{equation}
\begin{equation}
\kappa_{ps} = \frac{1}{T_p} \left(\Lfun{2}{ps} - \frac{\left[\Lfun{1}{ps}\right]^2}{\Lfun{0}{ps}}\right)\geq 0,
\label{eq:kappa_ps}
\end{equation}
respectively. $\kappa_{ps}\geq 0$ was proven in Ref.\ \cite{Shastry16}.

\subsection{Broad-band limit}

If the probe-sample coupling 
is broad-band, we may approximate $\Gamma^p(\omega)\approx \Gamma^p(\mu_0)$, 
where $\mu_0$ is the electrochemical potential of the source, drain, and probe
electrodes when the whole system is in equilibrium. Writing
\begin{equation}
{\rm Tr}\left\{\Gamma^p(\mu_0)\right\} = \bar{\Gamma}^p 
\label{eq:gamma_bb}
\end{equation}
and using Eq.\ (\ref{eq:Abar}), with $\Gamma^p(\omega)$ replaced by $\Gamma^p(\mu_0)$,
Eq.\ (\ref{eq:Ip_fs}) may be expressed as
\begin{equation}
I_p^{(\nu)}= \frac{\bar{\Gamma}^p}{\hbar} \int_{-\infty}^\infty\!\! d\omega (\omega-\mu_p)^\nu \,
\bar{A}(\omega) [f_s(\omega) -f_p(\omega)].
\label{eq:Ip_fs_bb}
\end{equation}

When the probe is in local equilibrium with the sample, as defined by Eq.\ (\ref{eq:def_probe}), 
Eqs.\ (\ref{eq:local_N}), (\ref{eq:local_E}), and (\ref{eq:Ip_fs_bb}) imply
\begin{eqnarray}
\left.\langle N\rangle\right|_{f_s} & = & \left.\langle N\rangle\right|_{\bar{f}_p}, 
\label{eq:n_ave_eq} \\
\left.\langle E \rangle\right|_{f_s} & = & \left.\langle E \rangle\right|_{\bar{f}_p}.
\label{eq:E_ave_eq}
\end{eqnarray}
That is to say, the mean local occupancy and energy of the nonequilibrium system are the same as if its local (nonequilibrium) spectrum
$\bar{A}(\omega)$ were populated by the equilibrium Fermi-Dirac distribution of the probe.
A non-invasive
measurement of $\bar{\mu}_p$, $\bar{T}_p$ thus constitutes a measurement of the local occupancy and energy of the system, in the broad-band limit. 
The quantity $\langle E\rangle -\bar{\mu}_p\langle N\rangle$ is a monotonically increasing function of $\bar{T}_p$ at fixed $\bar{\mu}_p$,
and is a measure of the mean excitation energy of the system (see Sec.\ \ref{sec:1st_law}).  Thus $\bar{T}_p$ {\em is directly related to the degree of
local energy excitation, in the same way as it is in an equilibrium system}.  Certainly, then, a floating thermoelectric probe provides more than a mere
operational definition of local temperature.

\begin{figure}[tb]
\begin{minipage}[c]{16cm}
        \includegraphics[width=10cm]{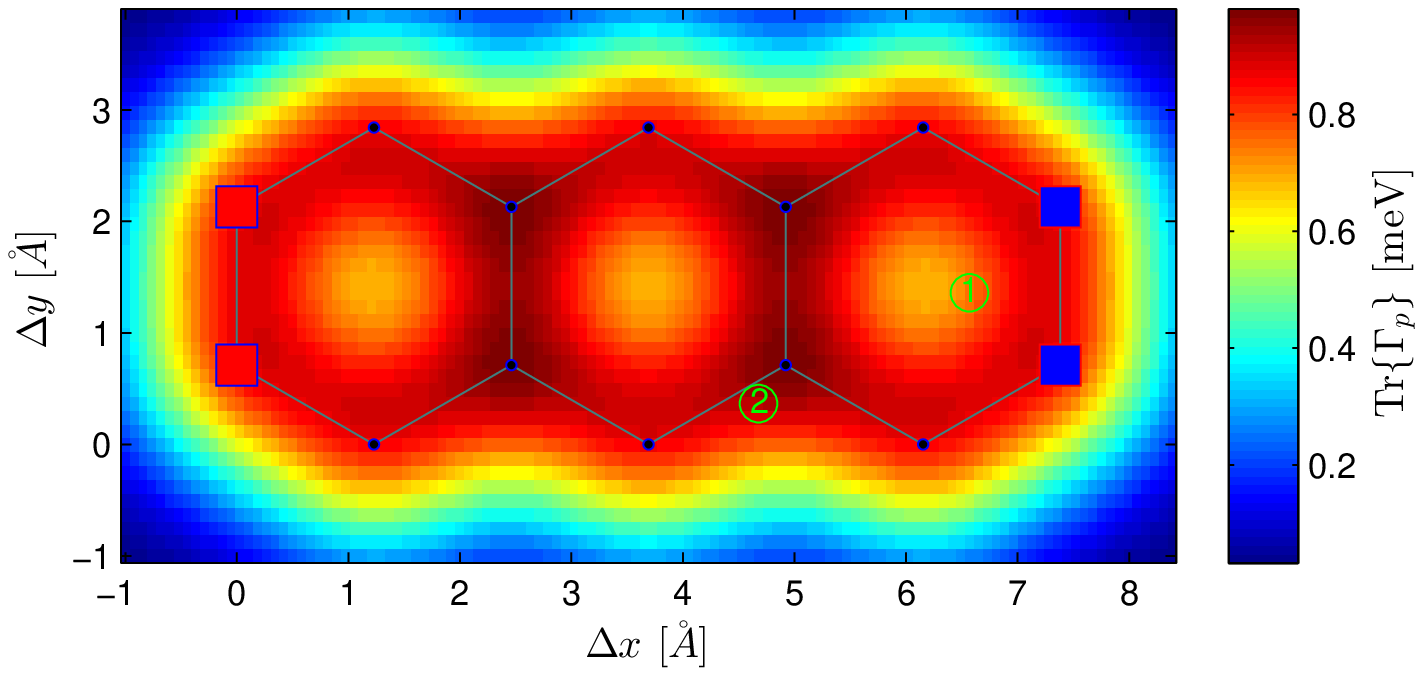}
\end{minipage}
\begin{minipage}[c]{8.5cm}
        \includegraphics[width=9.4cm]{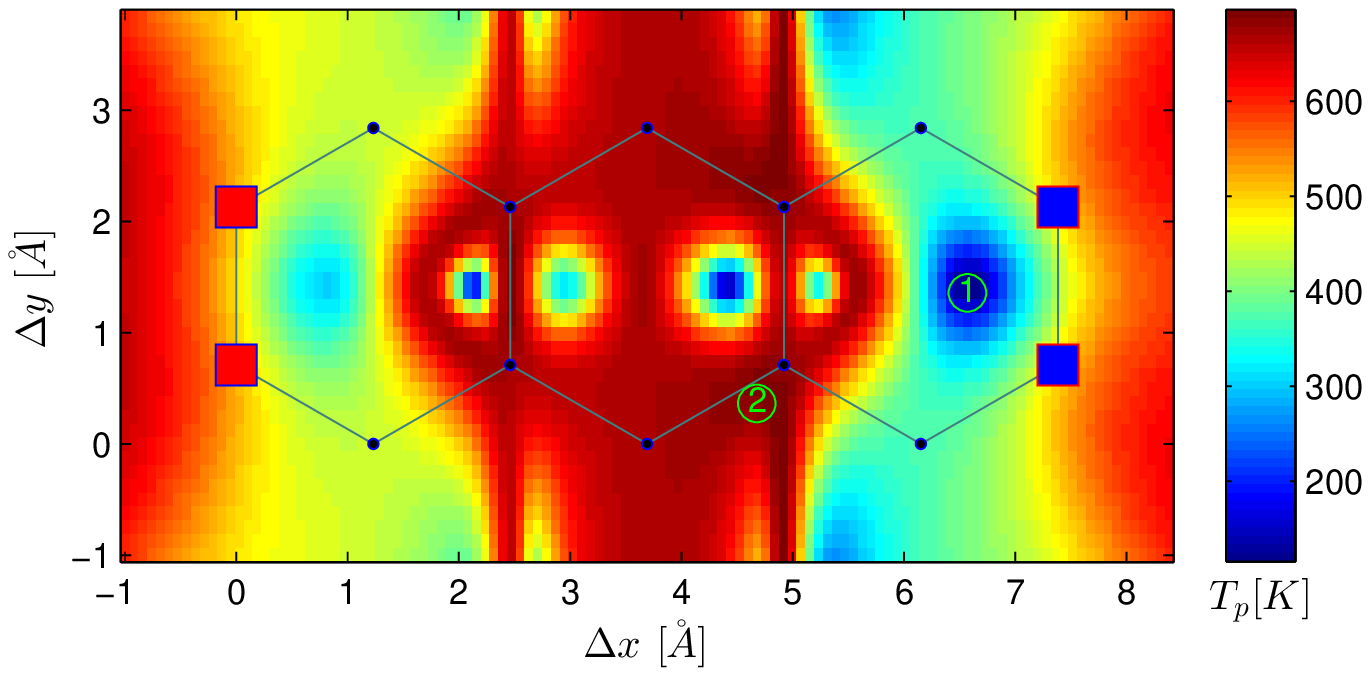}
\end{minipage}
\begin{minipage}[c]{8.5cm}
        \includegraphics[width=9.4cm]{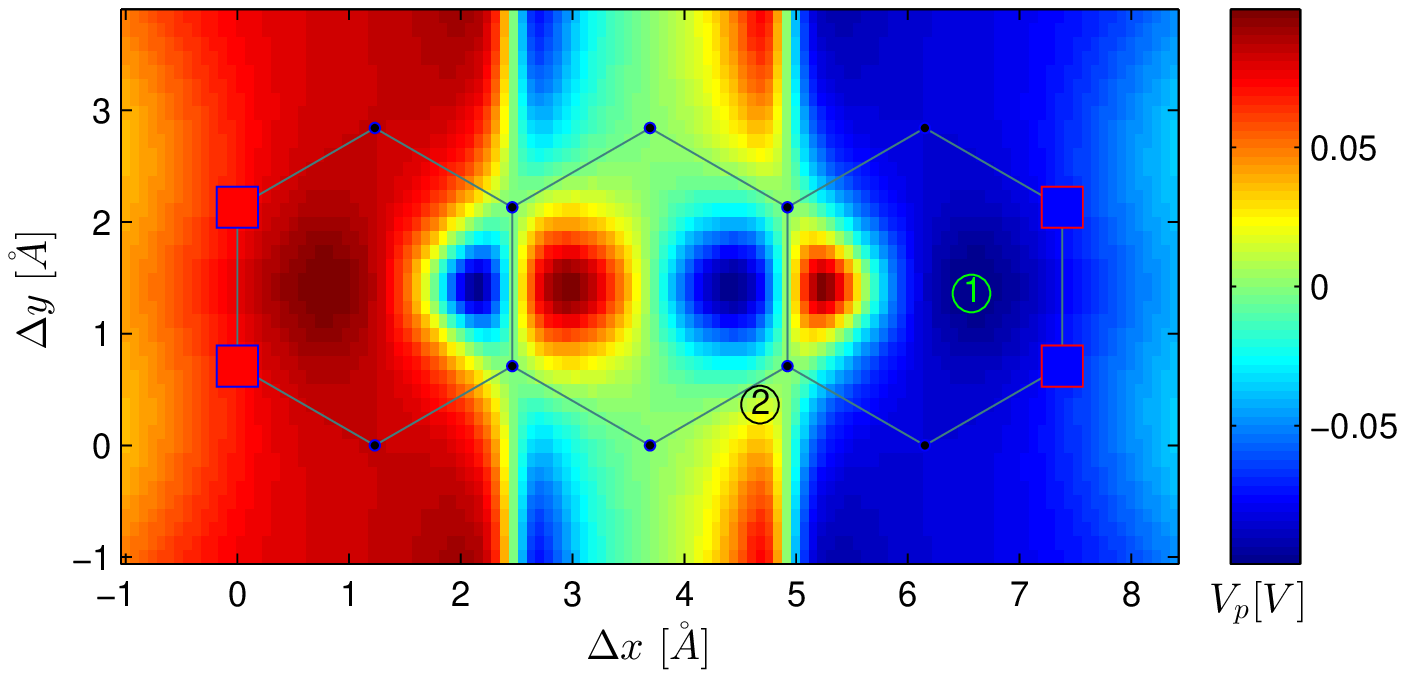}
\end{minipage}
\begin{minipage}[c]{8.5cm}
        \includegraphics[width=7.0cm]{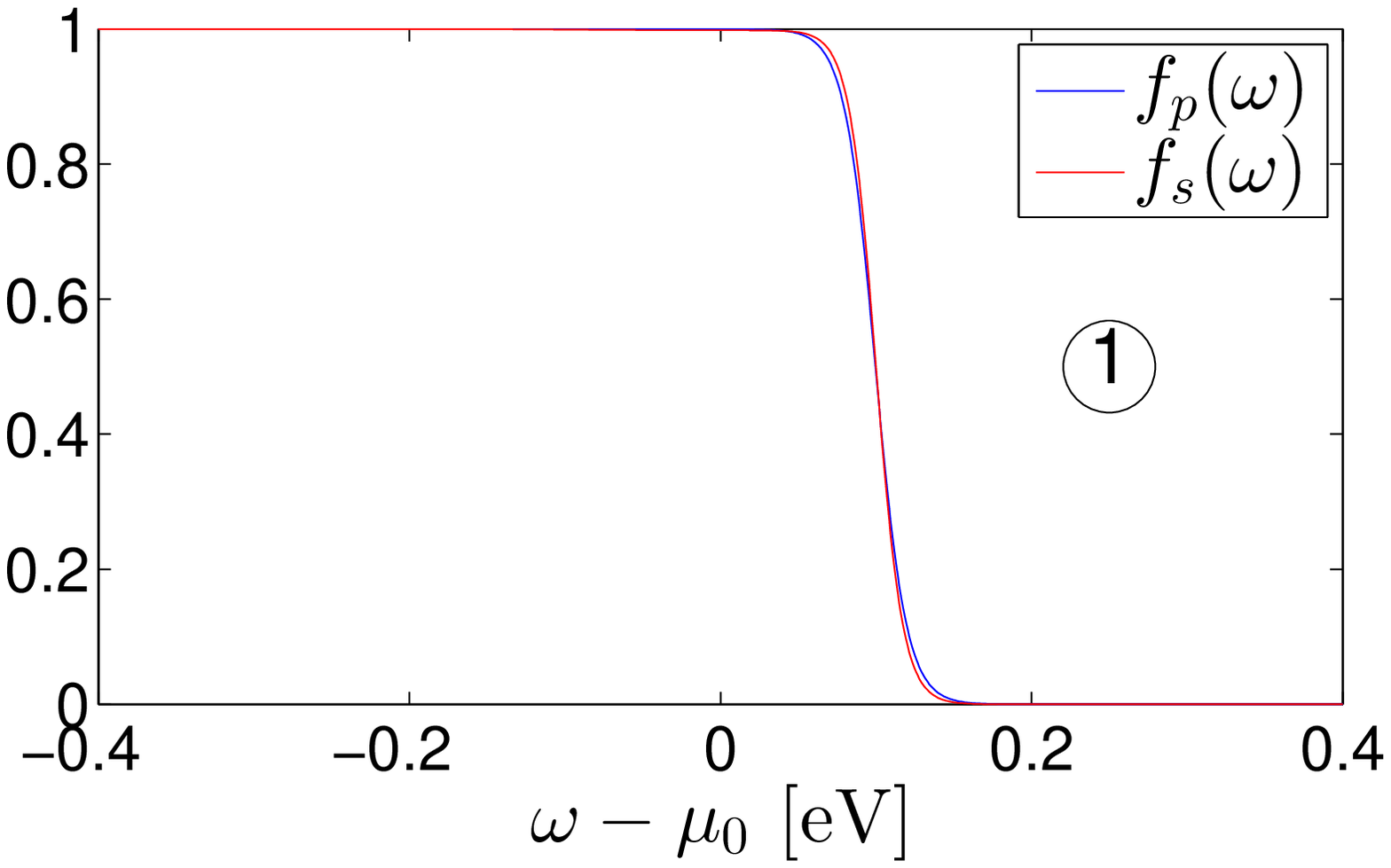}
\end{minipage}
\begin{minipage}[c]{8.5cm}
        \includegraphics[width=7.0cm]{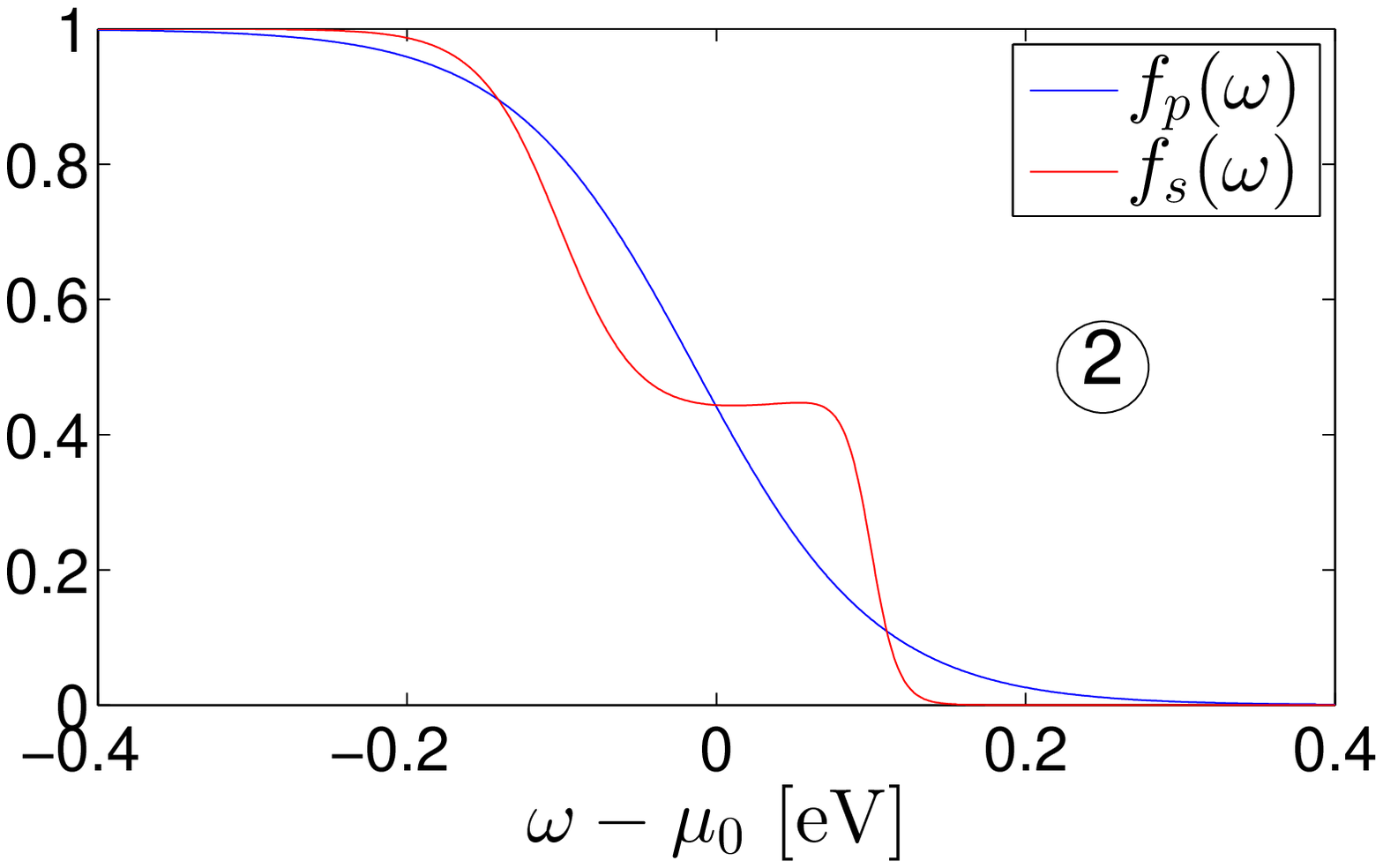}
\end{minipage}
\caption{%(Color online)
Measurements of a floating thermoelectric probe scanned 3.5\AA\
above the plane of carbon nuclei in a single-molecule junction containing an anthracene molecule. 
The electronic structure of the molecule is illustrated in the topmost panel, which shows
$\mbox{Tr}\{\Gamma^p(\mu_0)\}$ as a function of the probe's horizontal position.
The local temperature $\bar{T}_p$ and voltage $\bar{V}_p\equiv -\bar{\mu}_p/e$ are shown in the left and right panels of the middle row, respectively.
The probe is modeled as an atomically sharp Au tip and $\Gamma^p(\omega)$ was taken as a constant
evaluated at the Au Fermi energy (broad-band limit).  The thermoelectric bias of the junction is applied by two electrodes covalently bonded to the molecule
at the points labeled by the blue squares (electrode 1) and red squares (electrode 2), with $T_1=100\mbox{K}$, $T_2=300\mbox{K}$, and 
$\mu_2-\mu_1=0.2\mbox{eV}$.
The local energy distribution of the system $f_s(\omega)$ and the Fermi-Dirac distribution of the probe
$\bar{f}_p(\omega)$ are shown in the lower two panels for two different probe positions, corresponding to a cold spot and a hot spot, respectively
(indicated by circles in the top panels).  The zeroth and first moments of the probe's and system's local energy distributions are equal, as described by
Eqs.\ (\ref{eq:local_N})--(\ref{eq:local_E}) and (\ref{eq:n_ave_eq})--(\ref{eq:E_ave_eq}).
}
\label{fig:local_energy}
\end{figure}

The principle underlying Eqs.\ (\ref{eq:n_ave_eq})--(\ref{eq:E_ave_eq}), that a floating thermoelectric probe 
whose coupling to the system is broad band measures the zeroth and first moments of the system's local energy distribution,
is illustrated in Fig.\ \ref{fig:local_energy}.  For this example, the electronic transport was considered elastic, as described in Sec.\ \ref{sec:elastic}.
The electronic structure of the system (an anthracene molecule covalently bonded to two metal electrodes) 
was modeled via H\"uckel theory, and the floating thermoelectric probe was modeled as an atomically sharp Au tip 
scanned at a constant height of 3.5\AA\ above the plane of carbon nuclei in the junction.  The probe-system coupling was calculated by the method
of Refs.\ \cite{Bergfield13demon,Bergfield15}. The probe temperature $\bar{T}_p$ and chemical potential $\bar{\mu}_p$ were obtained by
finding the roots of Eq.\ (\ref{eq:both_currents}) numerically at finite bias \cite{Shastry15}.
At both the cold spot and hot spot indicated, the probe's Fermi-Dirac distribution matches the zeroth and first moments of the local energy distribution.
It should be emphasized that
this particular nanostructure is merely an example, chosen to illustrate the general principles involved in a scanning thermoelectric 
measurement, and the methods and approximations used to treat it in no way limit the applicability of the 
arguments given in the remainder of the paper.

\section{Zeroth Law}
\label{sec:0th_law}

In a previous article \cite{Meair14}, it was shown that the local temperature measured by a scanning thermoelectric probe is consistent with the 
{\em zeroth law of thermodynamics}, also known as the {\em transitive property of equilibrium}: if the local temperatures and chemical potentials of two 
nonequilibrium quantum systems, as measured by the probe, are equal, then the two systems will be in thermal and electrical
equilibrium with each other when connected by a 
transmission line coupled locally to the same two points.  This result was proven within linear response and to leading order in the Sommerfeld expansion.

This scenario can be extended to the nonlinear response regime, as discussed in Appendix \ref{sec:appendixB}.  However, here we focus on
another zeroth law scenario, namely: {\em Under what conditions will two different thermometers measure the same local temperature of a single nonequilibrium 
quantum system?} 

%\subsection{Broad-band limit}
\subsection{Ideal probe: local, non-invasive, broad-band}
\label{sec:zeroth_bbl}

%If the probe-sample coupling $\Gamma^p(\omega)$ is broad band, then the temperature and chemical potential of the probe are determined by
%Eqs.\ (\ref{eq:n_ave_eq}) and (\ref{eq:E_ave_eq}), and are thus unique, provided the local spectrum $\bar{A}(\omega)$ of the system is independent
%of the probe-sample coupling.  
It is well known in the field of scanning probe microscopy
\cite{Chen93} that the image of any physical property depends on the spatial resolution 
of the probe. 
This dependence drops out in
the limit of maximally local coupling given by Eq.\ (\ref{eq:Gamma_loc}), for which $\bar{A}(\omega)$ reduces to the local density of states
$\rho(\omega)$. 
For the sake of clarity, we focus on this limit of a locally coupled probe throughout the remainder of the paper. 
It is straightforward to extend these results to probes with arbitrary spatial resolution.

The local density of states $\rho(\omega)$ is independent of $\Gamma^p$ 
provided the coupling of the probe to the sample is not
so strong that it perturbs the local spectrum (non-invasive probe).  This may be seen explicitly as follows.  Using Dyson's equations for $G^r$ and $G^a$ 
\begin{equation}
G^{r/a} (\omega) = \left({\bf 1} \omega -H^{(1)} -\Sigma^{r/a}(\omega)\right)^{-1},
\label{eq:Dyson}
\end{equation}
where $H^{(1)}$ is the one-body Hamiltonian of the system and $\Sigma^{r/a}(\omega)$ is the retarded/advanced self-energy describing 
2-body interactions
and 
coupling of the system to the external reservoirs,
it can be shown that
\begin{equation}
A(\omega) = \frac{1}{2\pi} G^r(\omega) \Gamma(\omega) G^a(\omega),
\label{eq:spectrum2}
\end{equation}
where
\begin{equation}
\Gamma(\omega) \equiv -i\left(\Sigma^a -\Sigma^r\right) = \sum_\alpha \Gamma^\alpha -i \left(\Sigma_{\rm int}^a -\Sigma_{\rm int}^r\right).
\label{eq:Gamma}
\end{equation}
Here $\Gamma^\alpha(\omega)$ is the tunneling-width matrix describing coupling of reservoir $\alpha$ to the system, where $\alpha$ can represent source, 
drain, probe, etc., and $\Sigma_{\rm int}(\omega)$ is the self-energy due to two-body interactions.
Let $G_0^{r/a}=\lim_{\Gamma^p\rightarrow 0} G^{r/a}$ and $\Gamma_0 =\lim_{\Gamma^p\rightarrow 0} \Gamma$. Then $A_0(\omega) = G_0^r \Gamma_0 G_0^a$ is
the spectral function of the system in the absence of probe-system coupling, and one can show that
\begin{equation}
\frac{\rho(\omega)}{\rho_0(\omega)} = 1 - \frac{1}{2} \rho_0(\omega) \Gamma^p(\omega) + 
\frac{(G_0^r \Gamma^p G_0^a)_{nn}}{\rho_0(\omega)} 
+ {\cal O}\!\left(\rho_0 \Gamma^p\right)^2.
\label{eq:spectrum_correction}
\end{equation}
Any perturbation of the local spectrum by the probe can thus be safely neglected \cite{footnote_Gamma_size} provided $\rho_0(\omega) \Gamma^p \ll 1$.  
Similarly, the nonequilibrium distribution $f_s(\omega)$ is unaffected by the probe \cite{Shastry16}
provided $\Gamma^p \ll \Gamma^\alpha$ $\forall \alpha\neq p$, where
the $\alpha=1,2,\cdots$ denote the reservoirs of charge and energy used to drive the system out of equilibrium.

For any probe with such a maximally local, weak, broad-band coupling to the system, the measured value of the local temperature depends only
on the nonequilibrium state
of the system, and is independent of the properties of the probe.
The probe temperature and chemical potential are directly related to the mean energy and occupancy of the localized orbital to which it is coupled.
Any two such thermometers will measure exactly the same local temperature of the system, and thus satisfy the transitive property of equilibrium.  {\em The
local temperature so defined is thus consistent with the zeroth law of thermodynamics.}

The two conditions on the probe-system coupling needed to ensure consistency with the zeroth law, that it should be both weak and broad-band, are eminently 
reasonable, since they are needed to ensure that the measurement does not strongly perturb the system, and that the measurement depends on the
spectrum of the system rather than that of the thermometer, respectively.
We define such a measurement, where in addition the thermal coupling of the probe to the ambient environment is negligible ($\kappa_{p0}=0$), as an {\em ideal
temperature measurement}, and denote the value by $\hat{T}_p$.

\subsection{Beyond the broad-band limit}
\label{sec:zeroth_no_bbl}

To investigate deviations from the zeroth law far from equilibrium beyond the broad-band limit, one can solve Eqs.\ (\ref{eq:def_probe})
and (\ref{eq:both_currents}) for $\bar{T}_p$, treating $\Gamma^{p\prime}(\mu_0)$, $\Gamma^{p\prime\prime}(\mu_0)$, etc., as perturbations.
Let us define
\begin{equation}
\Gamma^p(\omega) \equiv \bar{\Gamma}^p [1+\lambda g(\omega)],
\label{eq:gdef}
\end{equation}
where $\lambda$ is a dimensionless parameter that is taken to be small and $g(\mu_0)=0$.
The temperature measured by the probe is
\begin{equation}
\bar{T}_p = \hat{T}_p + \delta T_p,
\label{eq:Tp_error}
\end{equation}
where $\hat{T}_p$ is the result for $\lambda=0$, and
it can be shown that the temperature error $\delta T_p$ of a nonideal thermometer with $\Gamma^p(\omega)\neq \mbox{const.}$ is
\begin{equation}
\delta T_p = \lambda \frac{\delta I_p^{(1)} + e \hat{T}_p S_{ps} \delta I_p^{(0)}}{\kappa_{ps}} + {\cal O}\left(\lambda^2\right),
\label{eq:delta_Tp}
\end{equation}
where 
$S_{ps}$ and $\kappa_{ps}$ are given by Eqs.\ (\ref{eq:S_ps}) and (\ref{eq:kappa_ps}), 
respectively, with Eq.\ (\ref{eq:Lnu_ps}) evaluated for 
$T_{ps}(\omega) \rightarrow 2\pi \bar{\Gamma}^p \rho(\omega)$ and 
$\bar{f}_p(\omega)\rightarrow \hat{f}_p(\omega)=f_p(\hat{\mu}_p,\hat{T}_p;\omega)$, where $\hat{\mu}_p\equiv \lim_{\lambda\rightarrow0} \bar{\mu}_p$ 
is the result of an {\em ideal voltage measurement}. 
Here 
\begin{equation}
\delta I_p^{(\nu)} =\frac{\bar{\Gamma}^p}{\hbar} \int_{-\infty}^\infty d\omega (\omega -\hat{\mu}_p)^\nu g(\omega) \rho(\omega) \left[f_s(\omega)-\hat{f}_p(\omega)\right].
\label{eq:delta_Ipnu}
\end{equation}

If one assumes that $g(\omega)$ is a slowly-varying function with
\begin{equation}
g(\omega) = g_1(\omega - \hat{\mu}_p) + g_2 (\omega-\hat{\mu}_p)^2 + \cdots,
\end{equation}
then one can show that 
\begin{equation}
\delta T_p = 
\frac{\lambda \bar{\Gamma}^p}{\hbar} 
\left(\frac{g_1 + e \hat{T}_p S_{ps} g_2}{\kappa_{ps}}\right)
\int_{-\infty}^{\infty} d\omega (\omega-\hat{\mu}_p)^2 \rho(\omega) [f_s(\omega) - \hat{f}_p(\omega)]
\label{eq:delta_Tp_2}
\end{equation}
plus corrections involving higher powers of $(\omega-\hat{\mu}_p)$ in the integrand.

In order to make further progress analytically, it is necessary to consider the limit of linear response.  
For 
small thermoelectric bias, transport in nanostructures is largely elastic at room temperature and below, so
one can use Eq.\ (\ref{eq:injectivity}) in Eq.\ (\ref{eq:delta_Tp_2}), and expand $f_\alpha(\omega)$ and $\hat{f}_p(\omega)$
about the equilibrium distribution $f_0(\omega)$.
We consider separately the cases of thermal and electrical bias.

\subsubsection{Thermal bias}
\label{sec:zeroth_thermal}

Evaluating Eq.\ (\ref{eq:delta_Tp_2}) for a thermal bias, and 
keeping only the leading term of the Sommerfeld expansion, one obtains
\begin{equation}
\delta T_p = \lambda g_1 \frac{7\pi^2}{5} \frac{\left(k_B T_0\right)^2}{\bar{T}_{ps}(\mu_0)} \sum_\alpha \bar{T}_{p\alpha}^{\, \prime}(\mu_0) (T_\alpha - T_p),
\label{eq:delta_Tp_th1}
\end{equation}
where 
\begin{equation}
\bar{T}_{p\alpha}(\omega)=2\pi \bar{\Gamma}^p \rho_\alpha(\omega)
\end{equation}
is the transmission probability from reservoir $\alpha$ into the probe,
evaluated in the broad-band limit for the probe-sample coupling.
One thus finds
\begin{equation}
\frac{\delta T_p}{\Delta T} 
\sim (k_B T_0)^2 
\left.\frac{d\ln\Gamma^p}{d\omega}\right|_{\omega=\mu_0}\!\!\! \left.\frac{d\ln \bar{T}_{p\alpha}}{d\omega}\right|_{\omega=\mu_0}.
\label{eq:delta_Tp_th2}
\end{equation}
for the temperature error as a fraction of the thermal bias.
The temperature error between any two thermometers is thus higher-order in the Sommerfeld expansion
for the case of a linear thermal bias,
and hence is expected to be numerically negligible for nanosystems at room temperature and below.
Note that the maximum temperature error between any two thermometers is bounded by $\Delta T$ for a pure thermal bias \cite{Bergfield13demon}, because
the equilibration of a thermometer at a
value of $T_p$ outside the range $[T_2,T_1]$ would violate the second law of thermodynamics.

\subsubsection{Electrical bias}
\label{sec:zeroth_electric}

Let us next consider a pure electrical bias, with both source and drain electrodes held at ambient temperature.  
Evaluating Eq.\ (\ref{eq:delta_Tp_2}), one obtains
\begin{equation}
\frac{\delta T_p}{T_0} \simeq \lambda g_1 \frac{7\pi^2}{10} \frac{\left(k_B T_0\right)^2 }{\bar{T}_{ps}(\mu_0)} \sum_\alpha \bar{T}_{p\alpha}^{\, \prime \prime}(\mu_0)
(\mu_\alpha - \hat{\mu}_p),
\label{eq:delta_Tp_el1}
\end{equation}
where the leading-order term of the Sommerfeld expansion vanishes due to condition (\ref{eq:def_probe}).
This temperature error should be compared to the temperature shift of an ideal probe due to the Peltier effect in the system \cite{Bergfield14_Buettiker}
\begin{equation}
\frac{T_p-T_0}{T_0} \simeq \frac {\bar{T}_{p1} \bar{T}_{p2}^{\, \prime} - \bar{T}_{p2} \bar{T}_{p1}^{\, \prime}}{\bar{T}_{ps}^2} \Delta \mu.
\label{eq:Tp_Peltier}
\end{equation}
The relative temperature error of a nonideal thermometer thus scales as
\begin{equation}
\frac{\delta T_p}{T_p-T_0} \sim \left(k_B T_0\right)^2 \left.\frac{d\ln\Gamma^p}{d\omega}\right|_{\omega=\mu_0} 
\frac{\bar{T}_{p\alpha}^{\, \prime \prime}(\mu_0)}{\bar{T}_{p\alpha}^{\, \prime}(\mu_0)}.
\label{eq:delta_Tp_el2} 
\end{equation}
As in the case of a thermal bias, the error is higher-order in the Sommerfeld expansion, and hence expected to be
numerically negligible for nanosystems at room temperature and below.

\section{First Law}
\label{sec:1st_law}

In this section, we investigate whether the temperature measured by a floating thermoelectric probe is consistent with the first law of 
thermodynamics.  We first consider a noninteracting system driven arbitrarily far from equilibrium, and show that the local temperature inferred from
an ideal temperature measurement is consistent with the first law.  We then consider an interacting system, where not only the local distribution
$f_s(\omega)$ but also the local density of states $\rho(\omega)$ depend on the nonequilibrium state of the system, and hence on the local 
temperature.  For this case, we show that deviations from the first law are higher-order in the Sommerfeld expansion.

\subsection{Noninteracting system}
\label{sec:1st_law_1body}

As in Sec.\ \ref{sec:zeroth_bbl}--\ref{sec:zeroth_no_bbl}, we focus here on the case of maximally local coupling given by Eq.\ (\ref{eq:Gamma_loc}), for which the 
quantities discussed in this section have an obvious meaning.  It is straightforward to generalize the arguments herein to arbitrary probe-sample coupling.

For a given nonequilibrium steady state of the system, the temperature of the probe is
determined by
\begin{equation}
0=I_p^{(1)}=\frac{\bar{\Gamma}^p}{\hbar} \int_{-\infty}^\infty d\omega \, (\omega -\bar{\mu}_p) \left[1+\lambda g(\omega)\right] \rho(\omega) 
\left[f_s(\omega) -\bar{f}_p(\omega)\right],
\label{eq:Ip1_1st_law}
\end{equation}
where we have used Eqs.\ (\ref{eq:Ip_fs}) and (\ref{eq:gdef}).  From Eqs.\ (\ref{eq:Ip1_1st_law}) 
and (\ref{eq:delta_Ipnu}), it follows immediately that
\begin{equation}
\left.\langle E \rangle\right|_{f_s} - \bar{\mu}_p \left.\langle N \rangle\right|_{f_s} =
\left.\langle E \rangle\right|_{\bar{f}_p} - \bar{\mu}_p \left.\langle N \rangle\right|_{\bar{f}_p} -
\lambda \frac{\hbar}{\bar{\Gamma}^p} \delta I_p^{(1)},
\label{eq:Q_fs}
\end{equation}
where $\langle E \rangle$ and $\langle N \rangle$ are the mean energy and occupancy, respectively, of the localized orbital of the system 
coupled to the probe, defined by
Eqs.\ (\ref{eq:local_N}) and (\ref{eq:local_E}).
To leading order in the Sommerfeld expansion, Eq.\ (\ref{eq:delta_Tp}) gives
\begin{equation}
\lambda \frac{\hbar}{\bar{\Gamma}^p} \delta I_p^{(1)}=\frac{\hbar}{\bar{\Gamma}^p} \kappa_{ps} \delta T_p
= C_s^{(1)}(\bar{\mu}_p,\bar{T}_p) \delta T_p,
\end{equation}
where the one-body contribution to the local specific heat is
\begin{equation}
C_s^{(1)}(\mu_p,T_p) = \frac{1}{T_p} \int_{-\infty}^\infty d\omega \, (\omega - \mu_p)^2 \rho (\omega) \left(-\frac{\partial f_p}{\partial \omega}\right)
\geq 0.
\label{eq:Cs1}
\end{equation}

Consider now a small change in bias of the nonequilibrium system, leading to a new nonequilibrium steady state characterized by the same value of the local
chemical potential $\bar{\mu}_p$, but by a different local temperature $\bar{T}_p^\prime$.  
The heat $\Delta Q_s$ added locally to the system under this change
of bias satisfies
\begin{equation}
\Delta Q_s \equiv \Delta\! \left\langle E -\bar{\mu}_p N \right\rangle = C_s^{(1)}(\bar{\mu}_p,\bar{T}_p)\, \Delta \! \left(\bar{T}_p - \delta T_p\right)
= C_s^{(1)}(\bar{\mu}_p,\bar{T}_p) \Delta \hat{T}_p,
\label{eq:dQs1}
\end{equation}
where Eqs.\ (\ref{eq:Q_fs})--(\ref{eq:Cs1}) have been used.
Here $\bar{T}_p - \delta T_p=\hat{T}_p$ is
the result of an {\em ideal temperature measurement} by a broad-band probe coupled weakly to the system, as discussed above in 
Sec.\ \ref{sec:0th_law}.  
Thus deviations from the zeroth and first laws under nonideal measurement conditions {\em are not independent}, and Eq.\ (\ref{eq:dQs1}) implies that it is
$\hat{T}_p$ that should be identified as the true local temperature of the system, directly related to the local energy excitation.

Note that Eq.\ (\ref{eq:dQs1}) also holds for systems with $\bar{T}_p <0$ (absolute negative temperature), 
%which may arise in strongly driven systems \cite{Shastry16}, 
although the interval
$[\bar{T}_p,\bar{T}_p^\prime]$ cannot contain 0
since $\langle E \rangle$ and $\bar{f}_p(\omega)$ are discontinuous at $\bar{T}_p =0$ (they are continous functions of $\bar{\beta}_p\equiv
1/k_B\bar{T}_p$).  Absolute negative temperatures do not
characterize any generic equilibrium state, but allow one to quantify population inversion in a nonequilibrium system with a bounded spectrum 
\cite{Ramsey56}. %,Braun13}.
Negative temperature solutions to Eq.\ (\ref{eq:def_probe}) exist for strongly driven systems \cite{Shastry16}.

\subsection{Interacting system}
\label{sec:1st_law_2body}

In an interacting system, not only the local distribution $f_s(\omega)$ but also the local spectrum $\rho(\omega)$ depends on temperature, so that
$ C_s = C_s^{(1)} + C_s^{(2)}$,
where
\begin{equation}
C_s^{(2)}(\mu_p, T_p)= \int_{-\infty}^\infty d\omega \, (\omega -\mu_p) f_p(\omega) \left.\frac{\partial \rho(\omega)}{\partial T_p}\right|_{\mu_p}
\label{eq:Cs2}
\end{equation}
is the two-body contribution to the local specific heat.
Due to the limited phase space for two-body scattering in Fermi systems \cite{AGD} 
at temperatures well below the Fermi temperature, 
$\partial \rho(\omega)/\partial T \propto T$ so that $C_s^{(2)}$ is two orders higher in the sense of a Sommerfeld expansion than $C_s^{(1)}$.

The arguments of Sec.\ \ref{sec:1st_law_1body} can be extended straightforwardly to the case of an interacting nonequilibrium system, leading to the result
\begin{equation}
\Delta Q_s \equiv \Delta\! \left\langle E -\bar{\mu}_p N \right\rangle 
= C_s^{(1)}(\bar{\mu}_p,\bar{T}_p) \Delta \hat{T}_p
+\int_{-\infty}^\infty d\omega \, (\omega -\bar{\mu}_p) \bar{f}_p(\omega) \Delta \rho(\omega),
\label{eq:dQs2}
\end{equation}
where $\Delta  \rho(\omega)$ is the change in the local spectrum due to the small change in bias of the nonequilibrium system.
For the case of a system driven out of equilibrium by a thermal bias alone, it is clear from the above discussion that the two-body term in Eq.\ (\ref{eq:dQs2}) is
two orders higher in the Sommerfeld expansion than the one-body term, and hence comparable to the error arising from a nonideal temperature measurement.
However, the size of the two-body term for general thermoelectric bias remains an open question.

Formally, one can write
$\Delta \rho(\omega) \equiv \Delta \hat{T}_p \partial \rho(\omega)/\partial \hat{T}_p$, where $\partial\rho(\omega)/\partial \hat{T}_p$ is the temperature derivative
of the spectrum of a fictitious equilibrium interacting system whose local spectrum coincides with that of the actual interacting nonequilibrium system.
In that case, of course, the first law applies also to the two-body contribution to $\Delta Q_s$, 
which characterizes the role of correlations in local heating of the nonequilibrium system.

\section{Second Law}
\label{sec:2nd_law}

In a previous article \cite{Meair14}, it was shown that if a nonequilibrium system is used as a heat bath to drive a thermoelectric process, the maximum
electrical work generated satisfies {\em Carnot's theorem}, with $\bar{T}_p$ as the absolute temperature of the bath.  That result was obtained within
linear response for noninteracting systems.
In this section, we demonstrate that the temperature measured by a floating thermoelectric probe satisfies Clausius' statement of the second
law of thermodynamics, that {\em no process is possible whose sole effect is to transfer heat from a system at some temperature $T$ to a system at a higher
temperature $T^\prime$}.  
The arguments of this section apply to steady-state systems arbitrarily far from equilibrium, and with arbitrary interactions.
The relation between probe temperature and the direction of heat flow was discussed in a different context by Caso {\it et al.} \cite{Caso11,Caso12}.

Let us consider the junction between the probe and the system. 
If the probe is biased 
away from the local equilibrium temperature $\bar{T}_p$ to some other temperature $T_p$, then a heat current $I_p^{(1)}$ will flow across the junction 
in accordance with Eq.\ (\ref{eq:Ip_2term}).  
It should be emphasized that $I_p^{(1)}$ is the {\em heat flowing into
the probe}, which is well defined, since the (macroscopic) probe is arbitrarily close to equilibrium in the presence of this microscopic heat current;
by contrast, the {\em heat flowing out of the system} is not well defined, since the system is far from equilibrium.

Eq.\ (\ref{eq:Ip_2term}) expresses the heat current $I_p^{(1)}$ in terms of the difference between two equilibrium
Fermi-Dirac distributions, $\bar{f}_p=f(\bar{\mu}_p,\bar{T}_p)$ and $f_p=f(\mu_p,T_p)$.  
$T_{ps}(\omega)$ in Eq.\ (\ref{eq:Ip_2term}) is given by
Eq.\ (\ref{eq:T_ps}) and satisfies $T_{ps}(\omega)\geq 0$ since both $\Gamma^p(\omega)$ and $A(\omega)$ are positive-definite. 
Thus Eq.\ (\ref{eq:T_ps}) gives the heat current across a fictitious two-terminal junction between two equilibrium reservoirs with transmission
function $T_{ps}(\omega)$.  

\subsection{Thermal bias of probe}
\label{sec:2nd_thermal_bias_only}

Let us first consider the case where the probe is thermally biased, but held at the equilibrium chemical potential $\bar{\mu}_p$.  Then 
$f_p(\omega) = f(\bar{\mu}_p,T_p; \omega)$, and
\begin{equation}
(\omega - \bar{\mu}_p) \left[\bar{f}_p(\omega)-f_p(\omega)\right] \stackrel{\textstyle >}{<} 0 \;\;\; \mbox{if} \;\;\; 
\beta_p \stackrel{\textstyle >}{<} \bar{\beta}_p,
\label{eq:heat_integrand}
\end{equation}
where $\beta_p \equiv 1/k_B T_p$.
Thus the integrand for $I_p^{(1)}$ in Eq.\ (\ref{eq:Ip_2term}) is everywhere positive for $\beta_p > \bar{\beta}_p$ and negative for 
$\beta_p < \bar{\beta}_p$,
so that 
\begin{equation}
{\rm sign}(I_p^{(1)}) = {\rm sign}(\beta_p-\bar{\beta}_p).
\label{eq:Clausius}
\end{equation}
That is to say, heat flows into the probe if it is biased to a temperature below the local temperature $\bar{T}_p$, and out of the probe for a bias above the
local temperature, consistent with the second law of thermodynamics.  This statement holds provided $T_p$ and $\bar{T}_p$ have the same sign; if 
$T_p>0$ and $\bar{T}_p<0$, then heat flows into the probe, in accordance with Eq.\ (\ref{eq:Clausius}) (a system at absolute negative temperature
is ``hotter'' than any positive temperature \cite{Ramsey56}), while the heat flow is reversed if the signs are
reversed.

This analysis rules out the possibility of multiple-valued solutions of
$I_p^{(1)}=0$
%Eq.\ (\ref{eq:def_probe}) 
at fixed $\mu_p=\bar{\mu}_p$.
The uniqueness of the probe temperature in the absence of electrical bias in the system (which precludes local Peltier cooling/heating effects)
was previously proven in Ref.\ \cite{Jacquet11}.

\subsection{Probe as open electric circuit}

Under the thermal bias conditions discussed above, a small electric current $I_p^{(0)}$ 
may flow across the junction between the probe and the system due to thermoelectric
effects.  To rigorously check the applicability of the Clausius formulation of the second law, we must consider 
a probe forming an open electric circuit, so that only heat may be exchanged between the probe and the system in steady state.
This leads to the condition $I_p^{(0)}=0$, 
which can be solved for the chemical potential shift $\Delta\mu_p=\mu_p-\bar{\mu}_p$ of the probe as a function of the thermal bias $T_p-\bar{T}_p$.  

For thermal biases achievable in the laboratory, the resulting thermoelectric voltage $\Delta\mu_p$ may be treated within linear response.
Writing 
\begin{equation}
\bar{f}_p \equiv f_p(\bar{\mu}_p,\bar{T}_p) = f_p(\mu_p,\bar{T}_p) + \left[\bar{f}_p - f_p(\mu_p,\bar{T}_p)\right] \cong 
f_p(\mu_p,\bar{T}_p) - \Delta\mu_p \left(\frac{\partial \bar{f}_p}{\partial \omega}\right),
\end{equation}
the open-circuit thermoelectric voltage may be obtained from Eq.\ (\ref{eq:Ip_2term}) as
\begin{equation}
{\cal L}^{(0)}_{ps} \Delta\mu_p \cong -\frac{1}{h} \int_{-\infty}^\infty d\omega \, T_{ps}(\omega) \left[f_p(\mu_p,\bar{T}_p)-f_p(\mu_p,T_p)\right],
\label{eq:Delta_mu_p}
\end{equation}
where ${\cal L}^{(0)}_{ps}$ is given by Eq.\ (\ref{eq:Lnu_ps}). 
The right-hand side of Eq.\ (\ref{eq:Delta_mu_p}) is just the electric current $I_p^{(0)}$ flowing when $\Delta\mu_p=0$.
The heat current flowing into the probe when it forms an open electric circuit may then be expressed as 
\begin{equation}
I_p^{(1)} \cong 
\left.I_p^{(1)}\right|_{\Delta\mu_p=0}
-\frac{{\cal L}^{(1)}_{ps}}{{\cal L}^{(0)}_{ps}} \left.I_p^{(0)}\right|_{\Delta\mu_p=0},
\label{eq:I_p1_open_circuit}
\end{equation}
where the first term is the heat current at $\Delta\mu_p=0$ 
discussed above, and the second term is a small thermoelectric correction, which has the 
opposite sign of the first term.  The thermoelectric correction is well known in the theory of electronic heat transport \cite{Onsager31,ZimanBook}. 
It represents
negative feedback arising from the ``interference'' of charge and heat transport processes \cite{Onsager31}, 
and cannot exceed the magnitude of the first term without leading to a violation of the second law.
Although the later condition has not been established in general for transport between an equilibrium system and a system far from equilibrium,
it must hold for the case at hand
due to the mapping onto a fictitious junction between two equilibrium systems provided by
Eq.\ (\ref{eq:Ip_2term}).

Thus, we have shown that the temperature measured by a floating thermoelectric probe satisfies Clausius' statement of the second law for arbitrary
steady-state
thermoelectric bias conditions of the system, and for arbitrary thermal bias between the probe and the system.
For a rigorous mathematical proof, see Ref.\ \cite{Shastry16}.

\section{Third Law}
\label{sec:3rd_law}

In this section, we investigate 
whether the local temperature of a nonequilibrium quantum system is consistent with the third law of thermodynamics.
From Eq.\ (\ref{eq:dQs2}), it follows that
\begin{equation}
\lim_{\bar{T}_p\rightarrow 0^+} \Delta Q_s \equiv \lim_{\bar{T}_p\rightarrow 0^+} \Delta\! \left\langle E -\bar{\mu}_p N \right\rangle
= C_s^{(1)}(\bar{\mu}_{p0},\bar{T}_p) \Delta \bar{T}_p,
\label{eq:dQs_3rd_law}
\end{equation}
where $C_s^{(1)}$ is given by Eq.\ (\ref{eq:Cs1}) and $\bar{\mu}_{p0}=\lim_{\bar{T}_p\rightarrow 0^+} \bar{\mu}_p$.  
Provided $\rho(\bar{\mu}_{p0}) \neq 0$, 
the low-temperature 
limit of $C_s^{(1)}$ may be straightforwardly calculated as
\begin{equation}
\lim_{\bar{T}_p\rightarrow 0^+} C_s^{(1)}(\bar{\mu}_p,\bar{T}_p) = \frac{\pi^2}{3} \rho (\bar{\mu}_{p0}) k_B^2 \bar{T}_p.
\label{eq:Cs1_lowT}
\end{equation}
Similarly, it can be shown that the leading-order behavior of the probe-sample thermal conductance $\kappa_{ps}$ defined in Eq.\ (\ref{eq:kappa_ps}) is
\begin{equation}
\lim_{\bar{T}_p\rightarrow 0^+} \kappa_{ps} = \frac{\pi^2 \bar{\Gamma}^p}{3\hbar} \rho (\bar{\mu}_{p0}) k_B^2 \bar{T}_p.
\label{eq:kappa_ps_lowT}
\end{equation}
Note that if $\rho(\bar{\mu}_{p0}) = 0$, both $C_s$ and $\kappa_{ps}$ vanish as higher powers of $\bar{T}_p$.
The fact that both $C_s \rightarrow 0$ and $\kappa_{ps}\rightarrow 0$ as $\bar{T}_p\rightarrow 0^+$ indicates that the local temperature inferred
from the measurement by a floating thermoelectric probe is {\em completely consistent with the third law of thermodynamics}.
Furthermore, it can be shown \cite{Shastry15} that the {\em local entropy of the system goes to zero} whenever $\bar{T}_p\rightarrow 0$.

Eqs.\ (\ref{eq:Cs1_lowT}) and (\ref{eq:kappa_ps_lowT}) also hold in the limit $\bar{T}_p\rightarrow 0^-$, with $\bar{T}_p$ replaced by $|\bar{T}_p|$.
These statements may be considered analogues of the third law \cite{Ramsey56} as it applies to the state of maximum energy in a system with a bounded spectrum.

\section{Conclusions}
\label{sec:conclusions}

In the present article,
a theory of local temperature measurement of an interacting quantum electron system arbitrarily far from equilibrium via a floating thermoelectric probe
was developed.  For a probe-system coupling that is both weak and broad-band, it was shown that the local temperature and chemical potential of
the probe are completely determined by the zeroth and first moments of the local energy distribution in the system [cf.\ Eqs.\ 
(\ref{eq:n_ave_eq})--(\ref{eq:E_ave_eq}) and Fig.\ \ref{fig:local_energy}].
The local temperature $\bar{T}_p$ so defined is thus directly related to the mean local excitation energy of the system
(\ref{eq:dQs1}), just as it is in an equilibrium
system.

%The local temperature so defined 
For a noninvasive broad-band probe,
it was shown that $\bar{T}_p$ is consistent with the zeroth, first, second, and third laws of thermodynamics.
For non-broad-band probes, %general probe-system couplings, 
the local temperature obeys the Clausius form of the second law
and the third law exactly, but there are deviations from the zeroth and first laws that are higher-order in the Sommerfeld expansion.
It was shown that the corrections to the zeroth and first laws are related, 
and can be interpreted in terms of the error inherent in a nonideal temperature measurement.
This analysis also applies to systems with negative absolute temperature \cite{Ramsey56,Braun13,Carr13,Shastry16} (population inversion).

The exact agreement with Clausius's statement of the second law and with the third law implies that the local temperature metric
$\bar{T}_p$ defines an ordering of temperatures, and an absolute zero, but not necessarily an absolute temperature scale.  The first law defines absolute
temperature differences, and it was shown that
discrepancies with the first law in probes that are not broad-band
arise from deviations from ideal measurement (zeroth law).
In this sense, {\em a noninvasive broad-band probe can be used to define an absolute temperature scale} for nonequilibrium quantum electron systems.  
All such 
thermometers will measure the same temperature, and the temperature scale so defined is consistent with the laws of thermodynamics, as elucidated above.
However, other types of thermometers \cite{Meair14,Giazotto06} may not yield precisely the same temperature, and the values of $\bar{T}_p$, $\bar{\mu}_p$
determined by a floating thermoelectric probe may not be consistent with alternative formulations of the
laws of thermodynamics, all of which are equivalent for equilibrium systems (see Appendix \ref{sec:appendixB} for a discussion of an alternative formulation
of the zeroth law). 

The ability to consistently define local thermodynamic
variables such as the temperature \cite{Meair14,Shastry15} or chemical potential \cite{Buttiker88,Buttiker89,Bergfield14_Buettiker} 
points to the possibilty of constructing a thermodynamic description---if only a partial one---of far-from-equilibrium quantum systems.

\begin{acknowledgments}
C.A.S.\ gratefully acknowledges discussions with Abhay Shastry, Justin Bergfield, Philippe Jacquod, and Jonathan Meair, and assistance in producing
Fig.\ \ref{fig:local_energy} from Abhay Shastry.
This work was supported by the U.S.\ Department of Energy, Office of Science grant no.\ DE-SC0006699.
\end{acknowledgments}

\appendix

\section{The nonequilibrium steady state}
\label{sec:AppA}

The nonequilibrium steady state is described by a density matrix $\hat{\rho}$ that is time-independent.
The expectation values of observables are given by their usual prescription in statistical physics, e.g.,
\begin{equation}
\langle\hat{Q}\rangle=\Tr{\hat{\rho}\hat{Q}}= \sum_{\mu,\nu}{\rho_{\mu\nu}\langle\nu|\hat{Q}|\mu\rangle}.
\end{equation}

The ``lesser" and ``greater" Green's functions \cite{Stefanucci13} used in the paper are defined as follows
\begin{equation}
G^{<}_{\alpha\beta}(t)\equiv i\langle {d^{\dagger}_{\beta}(0)}d_{\alpha}(t)\rangle,
\end{equation}
while its Hermitian conjugate is
\begin{equation}
G^{>}_{\alpha\beta}(t)\equiv -i\langle d_{\alpha}(t){d^{\dagger}_{\beta}(0)}\rangle,
\end{equation}
where
\begin{equation}
d_{\alpha}(t)= e^{i\frac{\hat{H}}{\hbar}t}d_{\alpha}(0)e^{-i\frac{\hat{H}}{\hbar}t}
\end{equation}
evolves according to the Heisenberg equation of motion for a system with Hamiltonian $\hat{H}$.
Here, $\alpha$, $\beta$ denote basis states in the 1-body Hilbert space of the system.

The spectral representation uses the eigenbasis of the Hamiltonian $\hat{H}|\nu\rangle=E_{\nu}|\nu\rangle$, where $\nu$ denotes a many-body energy
eigenstate.
One may write the ``lesser" Green's function as
\begin{equation}
\begin{aligned}
G^{<}_{\alpha\beta}(\omega)=2\pi i\sum_{\mu,\mu{'},\nu}\rho_{\mu\nu}&\langle\nu|d^{\dagger}_{\beta}|\mu{'} \rangle\langle\mu{'} |{d_{\alpha}}|\mu\rangle\\
&\times\delta\bigg(\omega-\frac{E_{\mu}-E_{\mu{'}}}{\hbar}\bigg),
\end{aligned}
\end{equation}
while the ``greater" Green's function becomes
\begin{equation}
\begin{aligned}
G^{>}_{\alpha\beta}(\omega)=-2\pi i\sum_{\mu,\mu{'},\nu}\rho_{\mu\nu}&\langle\nu|{d_{\alpha}}|\mu{'} \rangle\langle\mu{'} |d^{\dagger}_{\beta}|\mu\rangle\\
&\times\delta\bigg(\omega-\frac{E_{\mu{'}}-E_{\nu}}{\hbar}\bigg).
\end{aligned}
\end{equation}
The spectral function $A(\omega)$ is given by
\begin{equation}
A(\omega)\equiv \frac{1}{2\pi i}\bigg(G^{<}(\omega)-G^{>}(\omega)\bigg),
\end{equation}
and can be expressed in the spectral representation as
\begin{equation}
\label{SpectralFunction}
\begin{aligned}
A_{\alpha\beta}(\omega)=\sum_{\mu,\mu{'},\nu}\bigg[\rho_{\mu\nu}&\langle\nu|d^{\dagger}_{\beta}|\mu{'} \rangle\langle\mu{'} |{d_{\alpha}}|\mu\rangle\\
 +&  \rho_{\nu\mu{'}}\langle\mu{'}|{d_{\alpha}}|\mu \rangle\langle\mu |d^{\dagger}_{\beta}|\nu\rangle\bigg]\\
 & \times\delta\bigg(\omega-\frac{E_{\mu}-E_{\mu{'}}}{\hbar}\bigg).
\end{aligned}
\end{equation}
% We show here that $0\leq f_{s}(\omega)\leq1$ $\forall \omega\in(-\infty,\infty)$.

\subsection{Sum rule for the spectral function}
\label{sec:AppC}
Eq.\ (\ref{SpectralFunction}) leads to the following sum rule for the spectral function:
\begin{equation}
\label{SumRule}
\begin{aligned}
\int_{-\infty}^{\infty}d\omega A_{\alpha\beta}(\omega)=&\sum_{\mu,\nu}\rho_{\mu\nu}\langle\nu|d^{\dagger}_{\beta}d_{\alpha}|\mu\rangle\\
&\ \ \ \ \ \ \ \ \ \ \ \ \ \ +\sum_{\mu{'},\nu}\rho_{\nu\mu{'}}\langle\mu{'}|d_{\alpha}d^{\dagger}_{\beta}|\nu\rangle\\
=&\sum_{\mu,\nu}\rho_{\mu\nu}\langle\nu|d^{\dagger}_{\beta}d_{\alpha}+d_{\alpha}d^{\dagger}_{\beta}|\mu\rangle\\
%=&\sum_{\mu,\nu}\rho_{\mu\nu}\langle\nu|\delta_{\alpha\beta}|\mu\rangle\\
=&\sum_{\mu,\nu}\rho_{\mu\nu}\delta_{\mu\nu}\delta_{\alpha\beta}\\
=&\delta_{\alpha\beta}\Tr{\hat{\rho}}\\
=&\delta_{\alpha\beta}.
\end{aligned}
\end{equation}

In our theory of local thermodynamic measurements, the quantity of interest is the local spectrum of the system sampled by the probe $\bar{A}(\omega)$,
defined in Eq.\ (\ref{eq:Abar}). This obeys a further sum rule in the
broadband limit ({\em ideal probe}), discussed below.

\subsubsection{Local spectrum in the broadband limit}

The probe-system coupling is energy independent in the broadband limit, $\Gamma^{p}(\omega)=\text{const}$, and we write $\Tr{\Gamma^{p}}=\bar{\Gamma}^{p}$
for its trace.
The local spectrum sampled by the probe $\bar{A}(\omega)$ defined in Eq.\ (\ref{eq:Abar})
can be written in the broadband limit as
\begin{equation}
\bar{A}(\omega)=\frac{1}{\bar{\Gamma}^{p}}\sum_{\alpha,\beta}\langle\beta|\Gamma^{p}|\alpha\rangle A_{\alpha\beta}(\omega).
\end{equation}
In this limit, it obeys a further sum rule:
\begin{equation}
\begin{aligned}
\int_{-\infty}^{\infty}d\omega \bar{A}(\omega)&= \frac{1}{\bar{\Gamma}^{p}}\sum_{\alpha,\beta}\langle\beta|\Gamma^{p}|\alpha\rangle \int_{-\infty}^{\infty}d\omega A_{\alpha\beta}(\omega)\\
&=\frac{1}{\bar{\Gamma}^{p}}\sum_{\alpha,\beta}\langle\beta|\Gamma^{p}|\alpha\rangle\delta_{\alpha\beta}\\
&=1.
\end{aligned}
\label{eq:sumrulelocal}
\end{equation}
The broadband limit is special in that the measurement is determined by the local properties of the system itself, and is not influenced by the spectrum of
the probe.  In this limit, the local spectrum $\bar{A}(\omega)$ obeys the sum rule (\ref{eq:sumrulelocal}) since the probe samples the same
subsystem at all energies.  One should not expect such a local sum rule to hold outside the broadband limit, since the probe samples different subsystems
at different energies.

\subsection{Diagonality of $\hat{\rho}$}

We have, for any observable $\hat{Q}$,
\begin{equation}
\begin{aligned}
\langle\hat{Q}(t)\rangle&= \sum_{\mu,\nu} \rho_{\mu\nu}\langle\nu|\hat{Q}(t)|\mu\rangle\\
&=\sum_{\mu,\nu} \rho_{\mu\nu}\langle\nu|e^{i\frac{\hat{H}}{\hbar}t}\hat{Q}e^{-i\frac{\hat{H}}{\hbar}t}|\mu\rangle\\
&=\sum_{\mu,\nu} \rho_{\mu\nu}e^{-i\frac{E_{\mu}-E_{\nu}}{\hbar}t}\langle\nu|\hat{Q}|\mu\rangle.
\end{aligned}
\end{equation}

The system observables must be independent of time in steady state. Therefore $\hat{\rho}$ must be diagonal in the energy basis,
as seen from the above equation. The nondiagonal parts
of $\hat{\rho}$ in the energy basis, when they exist, must be in a degenerate subspace so that $E_{\mu}=E_{\nu}$ in the above equation.

For states degenerate in energy, the boundary conditions determining the nonequilibrium steady state will determine the basis in which $\hat{\rho}$
is diagonal. Henceforth, we work in that basis.

\subsection{Positivity of $-iG^{<}(\omega)$ and $iG^{>}(\omega)$}

Working in the energy eigenbasis in which $\hat{\rho}$ is diagonal, %we have
\begin{eqnarray}
\lefteqn{
-i\langle\alpha|G^{<}(\omega)|\alpha\rangle
 \equiv 
-iG^{<}_{\alpha\alpha}(\omega)  
=}
%\hspace*{-10mm}
\nonumber \\
 & & \!\!\!\!\!\!\! 2\pi\sum_{\mu,\mu{'}}\rho_{\mu\mu}\abs{\langle\mu|d^{\dagger}_{\alpha}|\mu{'}\rangle}^{2}
\delta\bigg(\omega-\frac{E_{\mu}-E_{\mu{'}}}{\hbar}\bigg) 
\geq 0.
\end{eqnarray}
Similarly,
\begin{eqnarray}
\lefteqn{i\langle\alpha|G^{>}(\omega)|\alpha\rangle \equiv
iG^{>}_{\alpha\alpha}(\omega) =} 
\nonumber \\
& & \!\!\!\!\!\!\! 2\pi\sum_{\mu,\mu{'}}\rho_{\mu\mu}\abs{\langle\mu|d^{\dagger}_{\alpha}|\mu{'}\rangle}^{2}
\delta\bigg(\omega-\frac{E_{\mu{'}}-E_{\mu}}{\hbar}\bigg)
\geq 0.
\end{eqnarray}
It follows that
\begin{equation}
\langle\alpha|A(\omega)|\alpha\rangle=
\frac{1}{2\pi}\langle\alpha|-\!iG^{<}(\omega)+iG^{>}(\omega)|\alpha\rangle
\geq 0.
\end{equation}
Therefore, all three operators $-iG^{<}(\omega)$, $iG^{>}(\omega)$, and $A(\omega)$ are positive-semidefinite.

\subsection{$0\leq f_{s}(\omega)\leq1$ and $\bar{A}(\omega)\geq 0$} %Boundedness of the nonequilibrium distribution function}
\label{Property:nonequilibriumdistribution}

The nonequilibrium distribution function $f_{s}(\omega)$ was defined in Eq.\ (\ref{eq:fs}) as
\begin{equation}
f_{s}(\omega)\equiv\frac{\Tr{\Gamma^{p}(\omega)G^{<}(\omega)}}{2{\pi}i\Tr{\Gamma^{p}(\omega)A(\omega)}}.
\label{eq:fs_appendix}
\end{equation}
We have $\Gamma^{p}(\omega)> 0$ by causality \cite{Stefanucci13}:
\begin{equation}
\text{Im}\ \Sigma^{r}_{p}(\omega)=-\frac{1}{2}\Gamma^{p}(\omega)<0.
\end{equation}
Let $\Gamma^{p}|\gamma_{p}\rangle=\gamma_{p}|\gamma_{p}\rangle$, where $\gamma_{p}\geq0$ and some $\gamma_{p}$ satisfy $\gamma_{p}>0$.
The energy dependence is taken to be implicit.
The traces in Eq.\ (\ref{eq:fs_appendix}) may be evaluated in the eigenbasis of $\Gamma^p$, yielding:
\begin{equation}
\begin{aligned}
f_{s}(\omega)&= \frac{\sum_{\gamma_{p}}\gamma_{p}\langle\gamma_{p}|G^{<}(\omega)|\gamma_{p}\rangle}{2\pi i\sum_{\gamma_{p}}\gamma_{p}\langle\gamma_{p}|A(\omega)|\gamma_{p}\rangle}\\
&= \frac{\sum_{\gamma_{p}}\gamma_{p}\langle\gamma_{p}|-i G^{<}(\omega)|\gamma_{p}\rangle}{\sum_{\gamma_{p}}\gamma_{p}\langle\gamma_{p}|-iG^{<}(\omega)+iG^{>}(\omega)|\gamma_{p}\rangle}.
\end{aligned}
\end{equation}
Therefore
\begin{equation}
0\leq f_{s}(\omega)\leq1.
\end{equation}
Similarly,
\begin{equation}
\begin{aligned}
\bar{A}(\omega)&\equiv \frac{\Tr{\Gamma^p(\omega)A(\omega)}}{\Tr{\Gamma^p(\omega)}} \\
&= \frac{\sum_{\gamma_{p}}\gamma_{p}\langle\gamma_{p}|-iG^{<}(\omega)+iG^{>}(\omega)|\gamma_{p}\rangle}{\sum_{\gamma_{p}}\gamma_{p}} \geq 0.
\end{aligned}
\end{equation}

\section{Alternative zeroth law scenario}
\label{sec:appendixB}

In this appendix, we consider the question investigated previously in Ref.\ \cite{Meair14}: If the local temperatures and chemical potentials of two
nonequilibrium quantum systems, as measured by a scanning thermoelectric probe, are equal, will the two systems be in thermal and electrical
equilibrium with each other when connected by a transmission line coupled locally to the same two points?  
This question was answered in the affirmative \cite{Meair14} for a noninteracting system 
within linear response and to leading order in the Sommerfeld expansion.  Here we extend the previous
analysis to consider two systems under arbitrary steady-state nonequilibrium conditions.

In this section, we consider noninteracting electrons and neglect the spin-orbit interaction, so we omit the spin index.
Let the fermion creation and annihilation operators of system $A$ be denoted by $d^\dagger$, $d$ and the corresponding Green's functions of system $A$
by $G^r$, $G^a$, and $G^<$, as defined in Sec.\ \ref{sec:intro_NEGF}.  Let the fermion creation and annihilation operators
in system $B$ be denoted by $f^\dagger$, $f$ and 
denote the retarded, advanced, and Keldysh ``lesser'' Green's functions describing
electron propagation/occupancy within system $B$ by
$g^r_{nm}(t)=-i\theta(t)\langle \{f_{n}(t),f_{m}^\dagger(0)\}\rangle$,
$g^a_{nm}(t)=i\theta(-t)\langle \{f_{n}(t),f_{m}^\dagger(0)\}\rangle$,
and $g^<_{nm}(t)=i\langle f_{m}^\dagger(0)\, f_{n}(t) \rangle$, respectively.

Suppose there is a point $a$ in system $A$ with local temperature $\bar{T}_p$ and chemical potential $\bar{\mu}_p$ as determined by a measurement
specified by Eqs.\ (\ref{eq:def_probe}) and (\ref{eq:Ip_fs}), and that there is a corresponding point $b$ in system $B$ characterized by the
same values of $\bar{T}_p$ and $\bar{\mu}_p$.  The question is whether points $a$ and $b$ will be in equilibrium with each other when connected
by a transmission line permitting the exchange of energy and charge.  

Let the Hamiltonian coupling systems $A$ and $B$ be 
\begin{equation}
H_{AB} = \sum_{\stackrel{\scriptstyle n\in a}{m\in b}} \left[V_{nm} d^\dagger_n f_m + \mbox{H.c.}\right].
\label{eq:H_AB}
\end{equation}
Then it can be shown using standard NEGF methods \cite{Meir92,Bergfield09a,Bergfield09b}
that the electric current $I_{AB}$ and energy current $I_{AB}^E$ flowing from system $B$ into system $A$ are given by
\begin{eqnarray}
I_{AB} & = & -\frac{e}{h} \int_{-\infty}^\infty d\omega \, {\rm Tr} \left\{\left[G^r(\omega)-G^a(\omega)\right]Vg^<(\omega)V^\dagger
+ G^<(\omega) V\left[g^a(\omega)-g^r(\omega)\right]V^\dagger\right\}, 
\label{eq:I_AB} \\
I_{AB}^E & = & \frac{1}{h} \int_{-\infty}^\infty d\omega \, \omega {\rm Tr} \left\{\left[G^r(\omega)-G^a(\omega)\right]Vg^<(\omega)V^\dagger
+ G^<(\omega) V\left[g^a(\omega)-g^r(\omega)\right]V^\dagger\right\}, 
\label{eq:I_AB^E}
\end{eqnarray}
respectively.  

For the case where $H_{AB}$ couples only a single localized orbital in system $A$ to a single localized orbital in system $B$
with matrix element $V$, Eqs.\ (\ref{eq:I_AB}) and (\ref{eq:I_AB^E}) can be simplified to
\begin{equation}
I_{AB}^{(\nu)} = \frac{2\pi |V|^2}{\hbar} \int_{-\infty}^\infty d\omega \, \omega^\nu \rho_a(\omega) \rho_b(\omega) \left[f_b(\omega)-f_a(\omega)\right],
\label{eq:I_AB^nu}
\end{equation}
where $\nu=0$ gives the fermion number current and $\nu=1$ gives the energy current.  
Here $f_a(\omega)$ and $f_b(\omega)$ are the local nonequilibrium distributions at points $a$ and $b$, respectively, defined according to
Eqs.\ (\ref{eq:fs}) and (\ref{eq:fn}), and $\rho_a(\omega)$ and $\rho_b(\omega)$ are the local densities of states at points $a$ and $b$,
respectively.

Notice that it is problematic in the present case to define
a heat current, since neither system $A$ nor system $B$ possesses a local equilibrium.  Nonetheless, the conditions 
\begin{equation}
I_{AB}^{(\nu)} = 0, \nu=0,1, 
\label{eq:AB_eq}
\end{equation}
suffice to define thermoelectric equilibrium between the two systems, and are equivalent to the conditions given by Eq.\ (\ref{eq:def_probe}) 
for the case where the heat current can be defined.

Eqs.\ (\ref{eq:def_probe}) and (\ref{eq:Ip_fs}) imply
\begin{eqnarray}
0 & = & \frac{1}{\hbar} \int_{-\infty}^\infty d\omega \; \omega^\nu \Gamma^p(\omega) \rho_a(\omega) \left[f_a(\omega)-f_p(\omega)\right], 
\;\;\; \nu=0,1,
\label{eq:IpA}
\\
0 & = & \frac{1}{\hbar} \int_{-\infty}^\infty d\omega \; \omega^\nu \Gamma^p(\omega) \rho_b(\omega) \left[f_b(\omega)-f_p(\omega)\right],
\;\;\; \nu=0,1.
\label{eq:IpB}
\end{eqnarray}
In linear response, it can be shown \cite{Meair14} that Eqs.\ (\ref{eq:IpA}) and (\ref{eq:IpB}) imply 
Eq.\ (\ref{eq:AB_eq})
to leading order in the Sommerfeld expansion.
Under general nonequilibrium conditions in systems $A$ and $B$, Eqs.\ (\ref{eq:IpA}) and (\ref{eq:IpB}) imply 
Eq.\ (\ref{eq:AB_eq}) provided $\Gamma^p(\omega)$, $\rho_a(\omega)$, and $\rho_b(\omega)$ can all be treated in the broad-band limit. That is to say, 
they can be taken as constant in the region where $f_b(\omega)-f_a(\omega)$ and $f_a(\omega)-f_p(\omega)$ differ significantly from zero.
Thus, the conditions for the validity of the zeroth law are somewhat more stringent for the scenario considered here than for the scenario considered
in Sec.\ \ref{sec:0th_law}.

\bibliography{refs}

\begin{thebibliography}{52}
\bibitem[{foo()}]{footnote_Gamma_size}
\bibinfo{note}{In our previous analysis \cite{Meair14} in the linear-response
  regime, it was found that the local temperature measured by the probe was
  essentially independent of the value of $\Gamma^p$ even for values as large
  as several eV, suggesting that the conditions on the strength of the
  probe-system coupling may be considerably less restrictive than those implied
  by Eq.\ (\ref{eq:spectrum_correction}).}
\end{thebibliography}

\end{document}